\documentclass[twocolumn]{aastex63}

\usepackage{ulem}

\shorttitle{Age Spreads and Systematics in $\lambda$ Orionis}
\shortauthors{Cao et al.}

\graphicspath{{./}{Figures/}}

\usepackage{graphicx}
\usepackage{amsmath}
\usepackage{amssymb}
\usepackage{tikz}

\newcommand{\lamOri}{$\lambda$ Orionis}
\newcommand{\lamori}{$\lambda$ Ori}
\newcommand{\vect}[1]{\mathbf{#1}}
\newcommand{\matr}[1]{\mathbf{#1}}

\begin{document}

\title{Age Spreads and Systematics in $\lambda$ Orionis with {\it Gaia} DR2 and the SPOTS tracks}

\correspondingauthor{Lyra Cao}
\email{cao.861@osu.edu, cao.lyra@gmail.com}

\author[0000-0002-8849-9816]{Lyra Cao}
\affiliation{Department of Astronomy, The Ohio State University, Columbus, OH 43210, USA}

\author[0000-0002-7549-7766]{Marc H. Pinsonneault}
\affiliation{Department of Astronomy, The Ohio State University, Columbus, OH 43210, USA}

\author{Lynne A. Hillenbrand}
\affiliation{Department of Astronomy, California Institute of Technology, Pasadena, CA 91125, USA}

\author[0000-0002-0631-7514]{Michael A. Kuhn}
\affiliation{Department of Astronomy, California Institute of Technology, Pasadena, CA 91125, USA}

\begin{abstract}
In this paper we investigate the robustness of age measurements, age spreads, and stellar models in young pre-main sequence stars. For this effort, we study a young cluster, $\lambda$ Orionis, within the Orion star-forming complex. We use {\it Gaia} data to derive a sample of 357 targets with spectroscopic temperatures from spectral types or from the automated spectroscopic pipeline in APOGEE Net. After accounting for systematic offsets between the spectral type and APOGEE temperature systems, the derived properties of stars on both systems are consistent. The complex ISM, with variable local extinction, motivates a star-by-star dereddening approach. We use a spectral energy distribution (SED) fitting method calibrated on open clusters for the Class III stars. For the Class II population, we use a {\it Gaia} G-RP dereddening method, minimizing systematics from disks, accretion, and other physics associated with youth. The cluster age is systematically different in models incorporating the structural impact of starspots or magnetic fields than in nonmagnetic models. Our mean ages range from 2-3 Myr (nonmagnetic models) to $3.9 \pm 0.2$ Myr in the SPOTS model ($f=0.34$). We find star-by-star dereddening methods distinguishing between pre-MS classes provide a smaller age spread than techniques using a uniform extinction, and infer a minimum age spread of $0.19$ dex and a typical age spread of $0.35$ dex after modelling age distributions convolved with observed errors. This suggests that the $\lambda$ Ori cluster may have a long star formation timescale and that spotted stellar models significantly change age estimates for young clusters.
\end{abstract}

\section{Introduction}
Stars form together in dense molecular cloud cores, in a process spanning orders of magnitude in the relevant timescales, length scales, and pressure scales. The age spread within a star cluster yields important clues about the star formation process, and a range of possibilities are viable \citep{2010ARA&A..48..581S,2014PhR...539...49K}.
Some authors suggest that stars must be formed within a cluster in a very small window of evolutionary time as a result of galactic dynamics and spiral arms \citep{2007RMxAA..43..123B,2007ApJ...668.1064E}. Others emphasize that pressure support from large-scale turbulence and magnetic fields can permit star formation over a more extended period of time \citep{2007ApJ...654..304K,2006ApJ...641L.121T}.
Pre-main sequence stars contract as they evolve, so age spreads should in principle be easily detected in star forming regions. However, even in cases where there is a large luminosity spread, such as in the Orion star-forming complex, it does not necessarily follow that a large age spread must be responsible \citep{1997AJ....113.1733H,2010ApJ...722.1092D}. Traditional stellar models neglect many of the physical processes important for young stellar objects, and evidence has been accumulating that these omissions impact our ability to understand young stars.

For example, the shapes of stellar isochrones can be inconsistent with the observed ones in star clusters, and this problem persists across multiple families of evolutionary models \citep{2010ARA&A..48..581S}. These distortions manifest as discrepancies in the ages inferred by low mass and high mass stars. There is a growing body of evidence that stellar magnetism and starspots may contribute to these isochronal age gradients. 
\citet{2016A&A...593A..99F} demonstrated that agreement between high and low-mass stars on the mass-radius and H-R diagrams is possible using magnetic models in Upper Sco. \citet{2019ApJ...884...42S} applied magnetic models to show that magnetic models produce masses in better agreement with data in Taurus and Ophiuchus. The first order effect of large spots is a distortion in the spectral energy distribution, as the emitted flux is not well-represented by a unique effective temperature model. However, starspots can also block convection and modify stellar structure. These effects can become very important when spots cover a large fraction of the surface, and there is good evidence that they do in young stars. \citet{2017ApJ...836..200G} presented a robust estimate of a very high spot filling fraction of 80\% for the weak-lined T Tauri star LkCa 4 in the similarly-aged Taurus Molecular Cloud.
\citet{2019ApJ...872..161D} applied a number of nonmagnetic, magnetic, and spotted tracks to Upper Sco eclipsing binary data, showing a consistent age for magnetic and spotted tracks on the mass-radius and the HR diagram. \citet{2020ApJ...891...29S} found that the mass-radius and HR diagram positions of eclipsing binaries in Upper Sco are better described by moderately spotted tracks. Forward modelling on the cluster, \citet{2020ApJ...891...29S} suggested a turnover in the isochronal age gradient between high mass and low mass stars occurring somewhere in a starspot filling fraction of $f_{\mathrm{spot}}=0.34$---$0.51$. Unresolved binaries may also perturb the isochronal age gradient \citep{2021ApJ...912..137S}.

To study the age spread and systematics in a young cluster, we choose the nearby ($\sim$400$\pm$40 pc) {\lamori} association (Collinder 69), a star forming region associated with the well-studied Orion star-forming complex. There are a variety of young stars spanning in mass from the brown dwarf limit to $\sim$24 $M_\odot$ historically associated with the cluster. It is an ideal laboratory for studying the age spread of young clusters because of its youth, proximity, and clear separation on the sky from other associations in Orion OB1a, with its kinematic separation making precise membership determination possible. A majority of stars associated with the {\lamOri} cluster show signs of H$\alpha$ or Ca II $8542$ emission and infrared excess \citep{2015AJ....150..100K}.
The {\lamori} cluster has both a center and a ring-like outer structure which has been suggested in the literature to have an elongated progenitor cloud \citep{lee2015,dolan2002}, with an associated supernova at its center. Many of its members have motions away from the center \citep{2018AJ....156...84K}, and the cluster itself appears kinematically unbound \citep{2019ApJ...870...32K}.
Molecular gas and photometric studies suggest an expanding ring of molecular clouds on the outskirts of the cluster \citep{maddalena1986,dolan2002}, and show that the structure has a nontrivial 3D structure with both spherical and ring-like components \citep{lee2015}. Wisps of this nebulosity, coupled with circumstellar material, result in non-uniform foreground extinction for each star, requiring individual extinction solutions for our sources, particularly in its kinematically distinct subregions B30 and B35.

Due to the significant variation in extinction between stars a consistent dereddening prescription is necessary in order to characterize the cluster.
In the literature some analyses of {\lamOri} have worked directly in color space, primarily to take advantage of the nearly parallel reddening vector in this cluster to avoid the effect of extinction altogether \citep{dolan2001}. In addition, techniques to estimate ages also have fit to a single cluster-wide value for extinction, as in the estimates of $5.6 ^{+0.4}_{-0.1}$ Myr with PARSEC isochrones assuming $A_V = 0.4$ \citep{2019A&A...628A.123Z}, 5---10 Myr \citep{2007ApJ...664..481B}, and 4 Myr \citep{1977MNRAS.181..657M}. The analysis from \citet{2018AJ....156...84K}, performs star-by-star dereddening with spectroscopic $T_{\mathrm{eff}}$'s, with an extensive kinematic analysis.

In this paper, we use isochrone ages with {\it Gaia} DR2 membership and apply a methodical treatment of extinction, membership, and temperatures to analyze stellar parameters with uncertainties. With a choice of an open-cluster calibrated methodology for addressing our Class II \& III sources, we incorporate both populations into our subsequent analysis. Classical and radius-inflated spotted or magnetic models are used to test the consistency between age estimates from the low and high mass stars. The resulting age distributions are tested, in different kinematically identified regions, and as a whole, to analyze the size of the luminosity or age spread.

In Section~\ref{sec:analysis} we discuss the methodology for inferring stellar parameters, including the photometric and spectroscopic data available for the cluster (Section~\ref{sec:data}), membership analysis with {\it Gaia} (Section~\ref{sec:membership}), classification of members into YSO class (Section~\ref{subsec:prefilter}), the intrinsic colors and reddening law (Section~\ref{sec:colortables}), our dereddening procedure (Section~\ref{sec:deredden}), SED fitting (Section~\ref{sec:sedfitting}), placement of stars on the HRD (Section~\ref{sec:put_hrd}), and age/mass estimation (Section~\ref{sec:massesages}). Results for {\lamOri} are presented in Section~\ref{sec:results}, including mean cluster age (Section~\ref{sec:clusterage}), age spread (Section~\ref{sec:agespread}), and systematic errors (Section~\ref{sec:systematics}) under multiple model assumptions.

\section{Analysis} \label{sec:analysis}
\begin{deluxetable*}{hcccccccccccccc}
\tabletypesize{\scriptsize}
\tablenum{1}
\tablecaption{Inferred stellar parameters with SPOTS $f_{\mathrm{spot}} = 0.34$ \label{tab:stellarparams}}
\tablewidth{0pt}
\tablehead{
\nocolhead{Star Name} &
\colhead{RA [deg]} &
\colhead{Dec [deg]} &
\colhead{$\mathrm{log \, T_{\mathrm{eff}}}$} &
\colhead{$\sigma _{\mathrm{log \, T_{\mathrm{eff}}}}$} &
\colhead{$\mathrm{log \, L/L_\odot }$} &
\colhead{$\sigma _{\mathrm{log \, L/L_\odot }}$} &
\colhead{$M/M_\odot$} &
\colhead{$\sigma _{M/M_\odot}$} &
\colhead{$\mathrm{log \, Age \, [yr]}$} &
\colhead{$\sigma _{\mathrm{log \, Age \, [yr]}}$} &
\colhead{$A_V$} &
\colhead{$\sigma_{A_V}$} &
\colhead{Class} &
\colhead{Region}
}
\startdata
\lbrack BNM2013 \rbrack 54.04   323&83.73008969&10.00963154&3.476&0.025&-1.124&0.302&0.166&0.052&6.575&0.306&0.351&0.678&III&LamOri-1\\
Haro 6-67&83.15163422&12.43873405&3.593&0.015&0.154&0.122&0.675&0.086&5.677&0.22&0.395&0.134&II&B30\\
\lbrack DM99 \rbrack  58&84.07932281&10.06413269&3.582&0.008&-0.398&0.07&0.739&0.083&6.62&0.151&0.135&0.16&III&LamOri-1\\
\lbrack BNM2013 \rbrack 53.02    97&83.84235182&9.89961398&3.5&0.028&-0.861&0.244&0.244&0.076&6.574&0.336&-0.287&0.586&III&LamOri-1\\
2MASS J05332402+1019535&83.35011292&10.33152866&3.622&0.008&-0.165&0.053&1.069&0.06&6.656&0.122&0.042&0.113&III&B30
\enddata
\tablecomments{Table 1 is published in its entirety in a machine-readable format.
      A portion is shown here for guidance regarding its form and content.}
\end{deluxetable*}

\subsection{Data} \label{sec:data}
\subsubsection{Photometry}

We aggregate optical, near-IR, and mid-IR photometry from a variety of literature sources for {\lamori} in a systematic manner.
This includes Gaia G, BP, and RP data \citep{2018A&A...616A...1G}, B, V, Rc, and Ic data from the AAVSO Photometric All Sky Survey (APASS) DR9 \citep{2016yCat.2336....0H} and a study of {\lamori} by \citet{dolan2001}, Pan-STARRS DR1 griz photometry \citep{flewelling2016}, and 2MASS J, H, and K$_{\mathrm{s}}$ data \citep{cutri2003}.
For mid-IR photometry, we crossmatch with WISE W1, W2, W3, and W4 photometry.

{\it Gaia} photometries (G, BP, RP) corresponding to apparent {\it Gaia} BP magnitudes fainter than 18.5 mag are discarded because of known spurious behavior in DR2 \citep{2018A&A...616A...1G}, and apparent Pan-STARRS DR1 photometry brighter than 14 mag as a result of published detector saturation thresholds \citep{2016arXiv161205560C}.

\subsubsection{Effective Temperature} \label{sec:temperature}

$T_{\mathrm{eff}}$ measurements are available for a variety of {\lamori} members. We aggregate spectral types from the literature, which includes work from \citet{2011A&A...536A..63B}, \citet{2012A&A...547A..80B}, \citet{2008A&A...488..167S}, \citet{dolan2001}, \citet{2015AJ....150..100K}, and \citet{2014yCat....102023S}. These collections of spectral types are included in the YSO Corral database. We convert these spectral types to $T_{\mathrm{eff}}$ with the empirical pre-MS relation from \citet{2013ApJS..208....9P}. The APOGEE Survey Data Release 14 \citep{2018ApJS..235...42A} released spectra for over 263,000 stars, including a large dataset within {\lamOri}. APOGEE Net \citep{2020AJ....159..182O,2018yCat..22360027C} provided a large, homogenous sample of spectroscopic effective temperatures for pre-MS stars, including a rich selection in {\lamori}. We crossmatch these stars with photometry from Pan-STARRS DR1 (griz) UCAC4 (B,V), 2MASS, and WISE.

\begin{figure}
\centering
\includegraphics[trim={0cm 0cm 0cm 0cm},clip,width=\columnwidth]{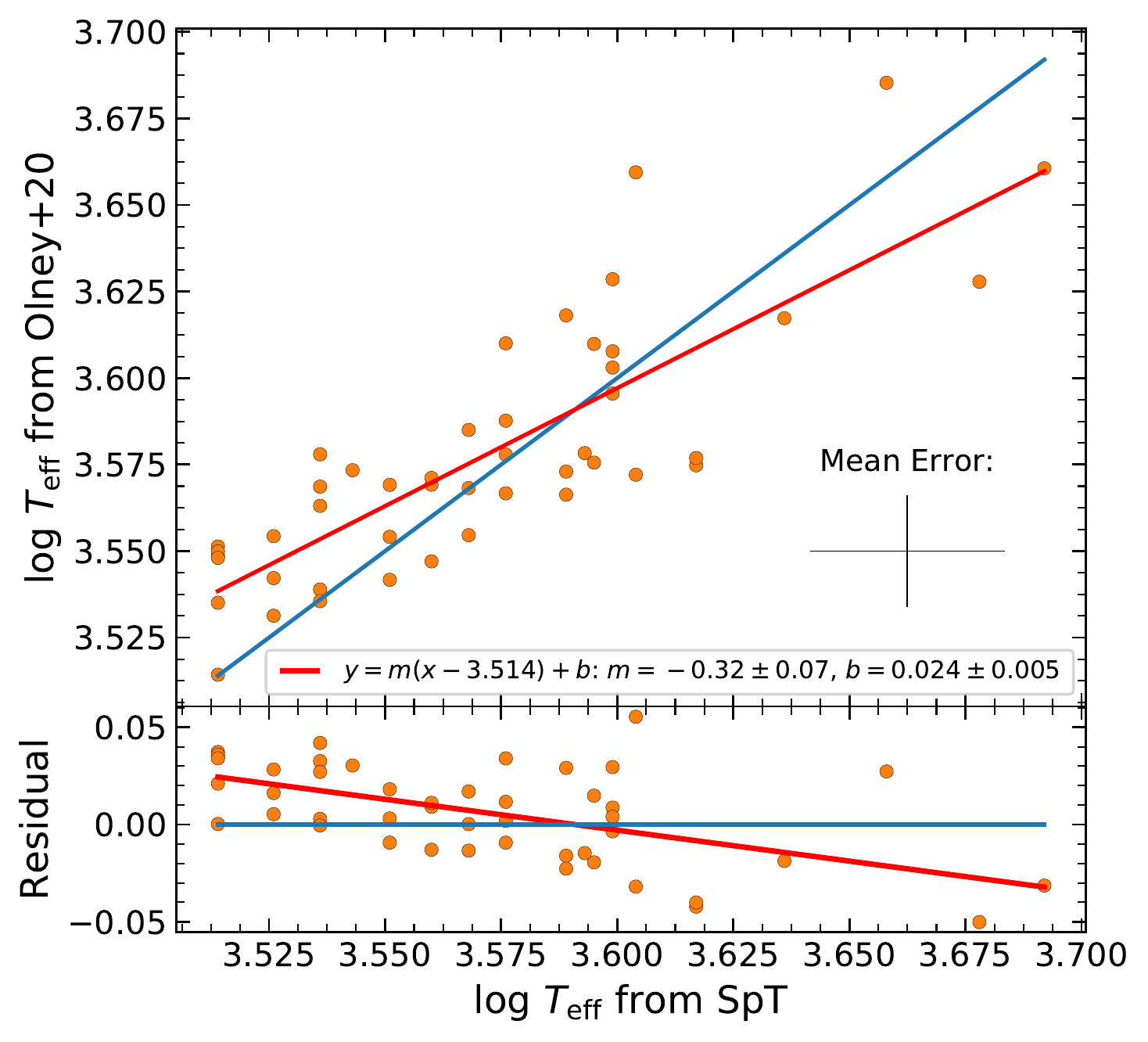}
\caption{Sample of stars common to our spectral type analysis and the \citet{2020AJ....159..182O} dataset. We observe a $>3 \sigma$ statistically significant nonzero slope and intercept in the data residuals, and correct for this temperature offset in logarithmic space in order to bring the APOGEE Net sample onto the same temperature system as the literature.} \label{fig:Figure_1}
\end{figure}

While the literature and APOGEE Net temperatures are both informed by fits involving spectra, the APOGEE Net pre-MS temperatures are generated from synthetic stars drawn from PARSEC isochrones. As a result, an offset between synthetic and literature temperatures is possible \citep[see discussion and Fig.2:][]{2020AJ....159..182O}. In our cross-sample between our sources and the APOGEE Net stars, we observe a trend in temperatures in Fig. \ref{fig:Figure_1}.
We apply a linear transformation in logarithmic temperature space to bring the APOGEE Net stars onto the same system as the traditional temperatures derived from spectral types.
As the rest of our analysis depends on work calibrated on pre-MS temperatures from the literature, we adopt this offset for all of the APOGEE Net stars in our sample.

\begin{figure}
\centering
\includegraphics[trim={0cm 0cm 1cm 1cm},clip,width=\columnwidth]{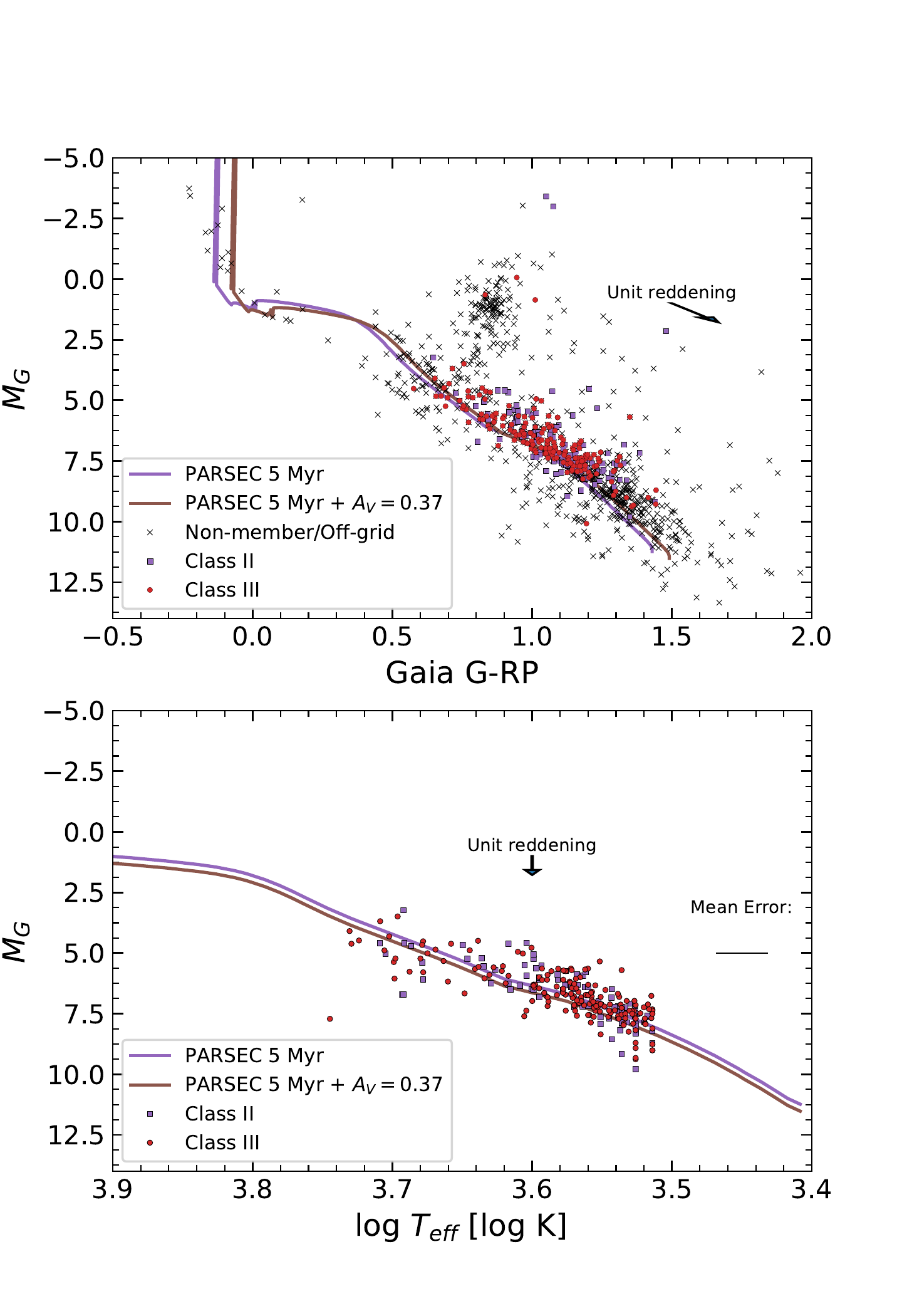}
\caption{Top: {\lamori} cluster CMD in {\it Gaia} G-band, {\it Gaia} G-RP space, with a PARSEC 5 Myr isochrone with a reddening of $0.37$ mag applied.
Open symbols are cluster members for which we have {\it Gaia} photometric and kinematic data.
Closed symbols are a subset for which we have independent spectral types for our subsequent analysis.
Bottom: absolute G-band magnitudes against inferred $T_{\mathrm{eff}}$.} \label{fig:Figure_2}
\end{figure}

Fig. \ref{fig:Figure_2} shows the cluster members in a CMD, with the top plot showing the wide number of cluster members for which we have photometric and kinematic data (in open symbols), and the ones that have independent spectroscopic $T_{\mathrm{eff}}$'s in the range covered by our color tables (in closed symbols).
Stars with and without significant active accretion (Class II and III respectively) are shown with purple and red symbols respectively; they are distinguished based on their far IR SED (Sec. \ref{subsec:prefilter}).

\subsection{Membership} \label{sec:membership}

{\it Gaia} data has made inferring cluster membership much more accurate \citep{2020MNRAS.496.4701J}.
Analyses of {\lamori} membership have historically been limited by the lack of accurate parallax data, despite the on-sky separation between {\lamOri} and the rest of the Orion star-forming complex.
Additionally, {\it Gaia} and APOGEE \citep{2018yCat..22360027C} have significantly increased the number of stars that are accessible with isochrone-based age-dating methods.
We obtain candidate \lamori\ members from published lists compiled in the YSO Corral database \citep{hillenbrand2021csss} and from APOGEE Net \citep{2020AJ....159..182O, 2018yCat..22360027C}. The combined list includes 1824 objects.

\begin{figure}
\centering
\includegraphics[trim={0.5cm 0cm 0cm 0cm},clip,width=\columnwidth]{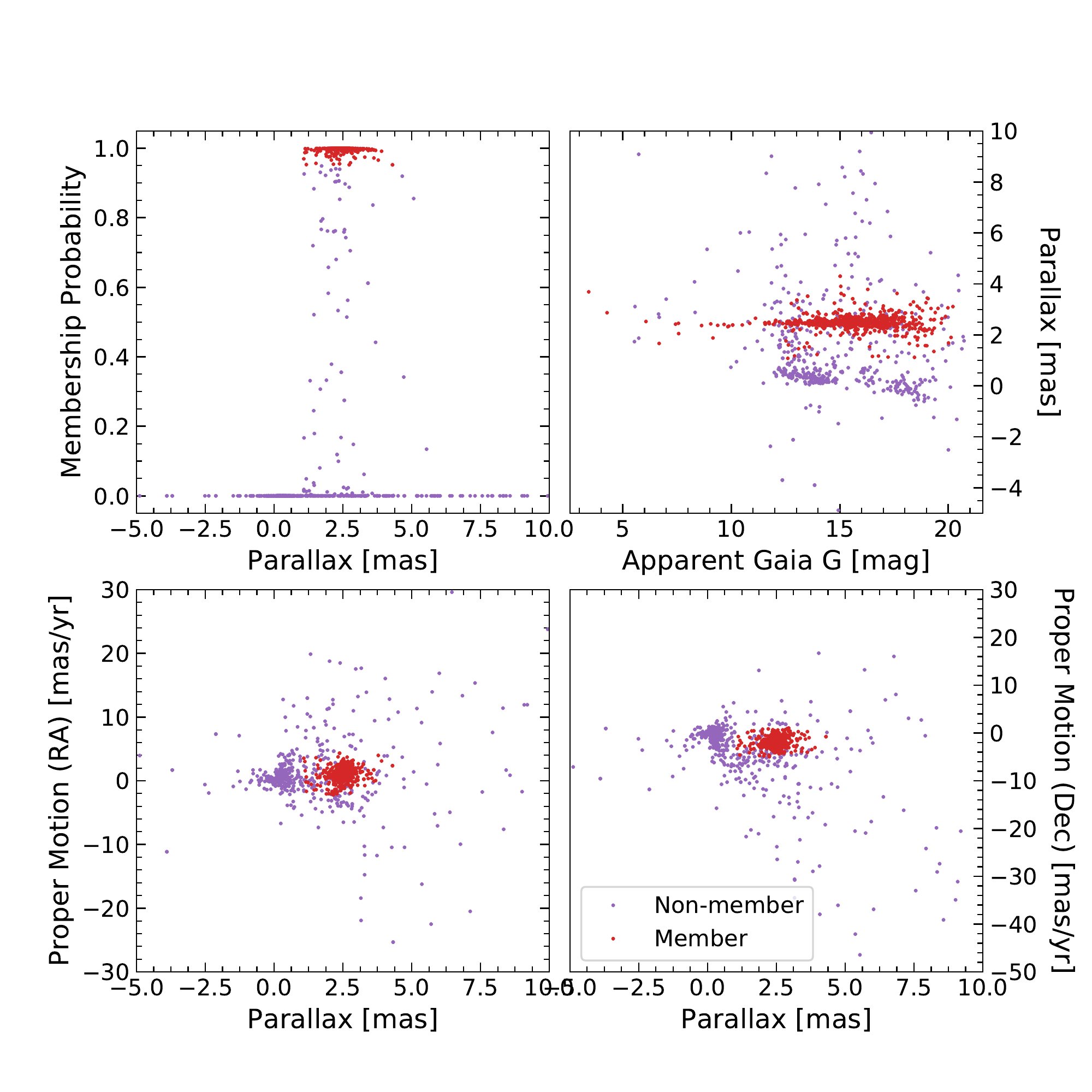}
\caption{The results of our membership analysis with our Gaussian Mixture Model. Top left: membership probability as a function of parallax. Top right: Members highlighted against field in apparent {\it Gaia} G-mag v. parallax. Bottom: Members separated in parallax--proper motion space.} \label{fig:Figure_3}
\end{figure}

Astrometric measurements from {\it Gaia} DR2 directly constrain membership in our two-component Gaussian Mixture Model. We assign membership probabilities to each star based on {\it Gaia} parallax and proper motion using the \textsc{scikit-learn} mixture model implementation \citep{scikit-learn}---with a normalization from \citet{2018PASP..130l4101G}. The code estimates the cluster center in parallax and proper motion using the expectation-maximization algorithm, where one Gaussian component represents the cluster and another the field star population, and iterates until convergence.
Out of 1824 candidate members, 1361 have applicable parallax information from {\it Gaia} DR2, and each of these stars has proper motion data.
We also apply a constraint on parallaxes to exclude stars which are systematically too far away. Since the parallax of the cluster center is $\sim$2.5~mas, we exclude stars that have parallaxes less than or equal to 1~mas. The identified cluster center at $\sim$2.5~mas has a dispersion of 0.3~mas, meaning the cut at 1~mas is at 4.8~$\sigma$. Fig. \ref{fig:Figure_3} demonstrates that we effectively distinguish between the cluster center and the field, identifying 875 stars as probable members at $P > 0.95$ and 486 as possible contaminants. The top-right panel shows that despite not considering photometry in our fit, we identify our cluster center in parallax across a wide range of apparent G-mag.

With a population positively identified as members of {\lamOri}, we distinguish between kinematically-distinct subregions. An additional membership analysis step involving RA/Dec is done to separate the cluster center with sources from B30 and B35. This is a four-dimensional fit across proper motion and RA/Dec with the same Gaussian Mixture Model. We tag these stars by the results of this kinematic analysis.

Our method for inferring $L$ and $T_{\mathrm{eff}}$ requires independent spectroscopic data, and we therefore limit our sample by requiring either a spectroscopic effective temperature or a spectral type. In our sample, 357 out of 875 probable members have spectral types or spectroscopic effective temperatures, including 140 from YSOC and 261 from APOGEE Net. For the 44 stars with temperatures from both sources, we choose to prioritize the spectral types reported from YSOC.

The stars in {\lamori} with available spectral types that also pass the astrometric membership criteria are predominantly low-mass FGKM stars, which is consistent with the predominantly K and M dwarfs available in previous studies \citep{2010ApJ...722.1226H}.

\subsection{Member Classification} \label{subsec:prefilter}
Many {\lamOri} members have disks, which can produce infrared excesses. For pre-main sequence stars with associated local extinction---especially the most shrouded, disky, and accreting sources---the reddest photometric bands ($\lambda > 1.25$ $\mu$m) can have reprocessed dust emission that is not accounted for in the dereddening prescription. Since we do not attempt to do modelling of the IR excess, nor the UV---blue excess due to accretion, we avoid using these bands in stars that are designated as Class II, which we define as systems with evidence of substantial reprocessed disk emission.

\begin{figure}
\centering
\includegraphics[trim={0cm 0cm 0cm 0cm},clip,width=\columnwidth]{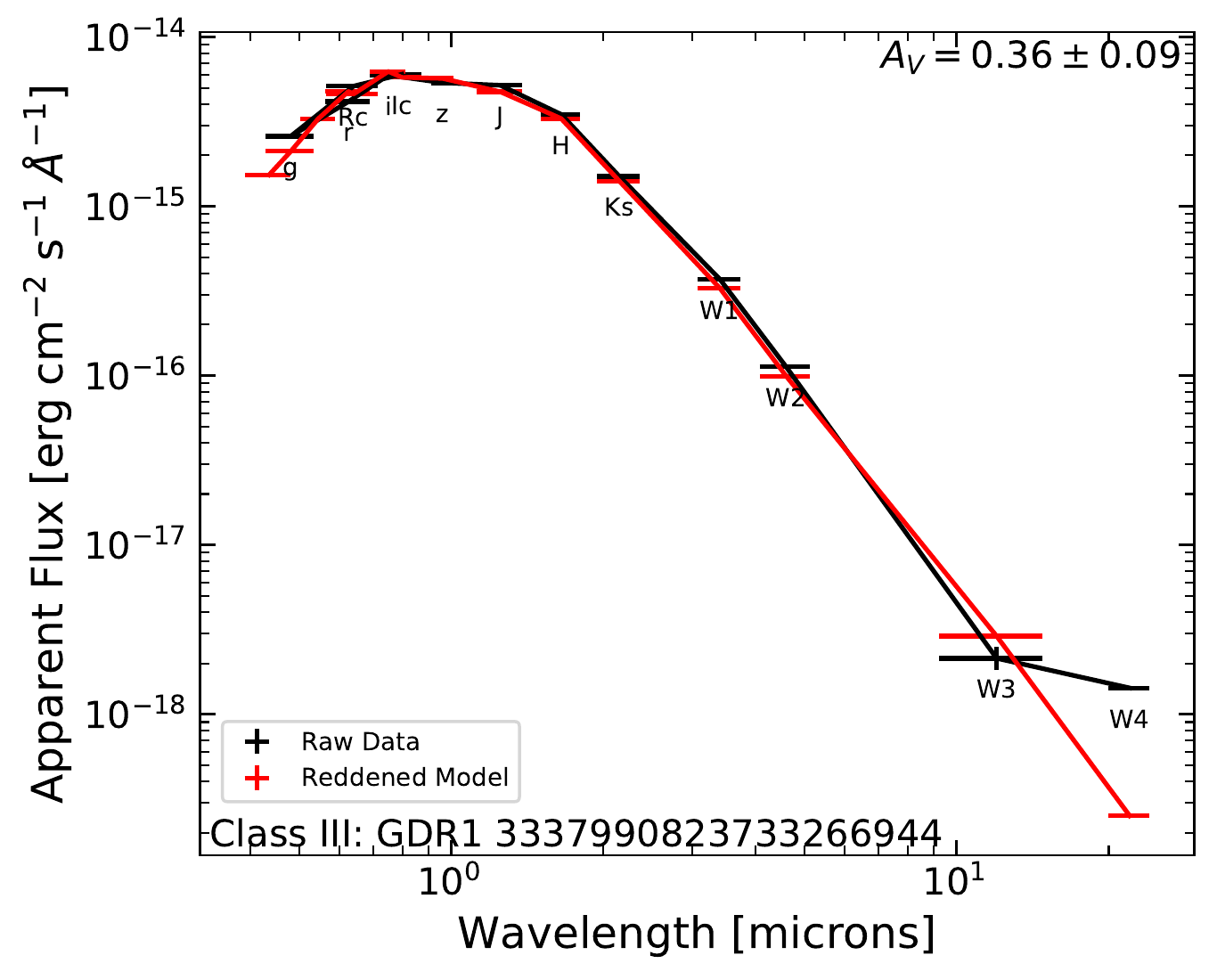}
\includegraphics[trim={0cm 0cm 0cm 0cm},clip,width=\columnwidth]{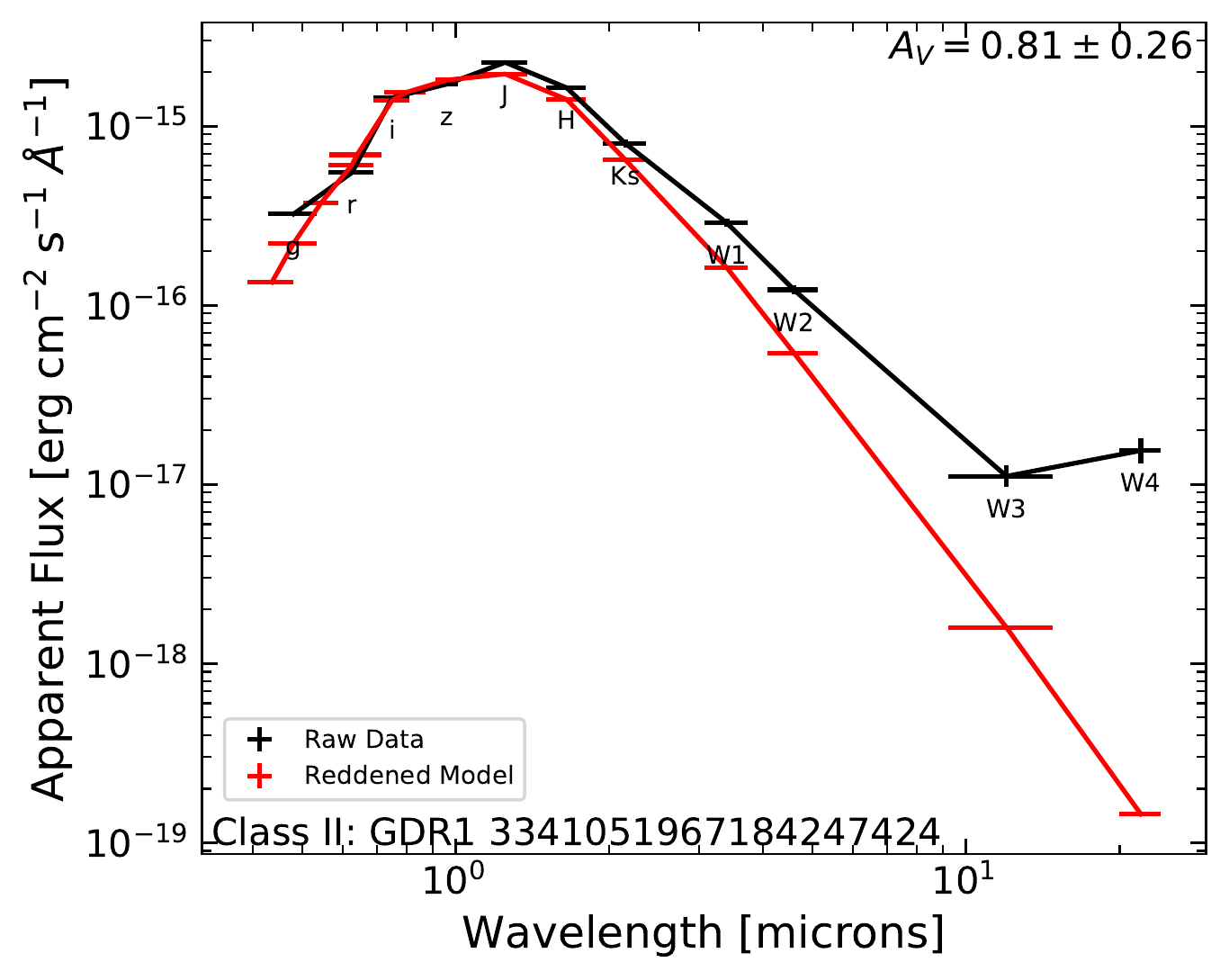}
\caption{A Class III and a Class II SED, both fit with our models from Sec. \ref{sec:colortables}. Top: the Class III star has little discrepancy between the template colors and the measured SED at all wavelengths. Bottom: the Class II star has a significant observed excess in the near-IR. This discrepancy is due to disks, circumstellar material, and other young star physics.} \label{fig:Figure_4}
\end{figure}

For Class II stars, we adopt
the WISE photometric criterion W1$-$W2$> 0.3$ (corresponding to SED slope $\alpha \ge -1.6$), which indicates that their SEDs have significant infrared excesses \citep{2015AJ....150..100K}. These IR excesses may be the result of reprocessed starlight, which our extinction model does not attempt to address.
The observed IR excess for these Class II stars is consistent with the comparable contributions from the photosphere and the disk in the near-IR \citep{2015PhDT.......219D}. Fig. \ref{fig:Figure_4} demonstrates a typical case of a fit to a Class III and a Class II star using a young star template.

With a fit from W1-W2 color to SED slope ($\alpha$) from \citet{2015AJ....150..100K}, we divide our population into Class II and Class III categories \citep[see][]{2019MNRAS.486.1907J}, a designation that we will use throughout this work:
\begin{equation*}
\begin{cases} \mathrm{W1-W2} < 0.3 &\to \mathrm{Class \, III}, \\
\mathrm{W1-W2} > 0.3 &\to \mathrm{Class \, II}.
\end{cases}
\end{equation*}
\begin{figure}
\centering
\includegraphics[trim={0cm 0cm 0cm 0cm},clip,width=\columnwidth]{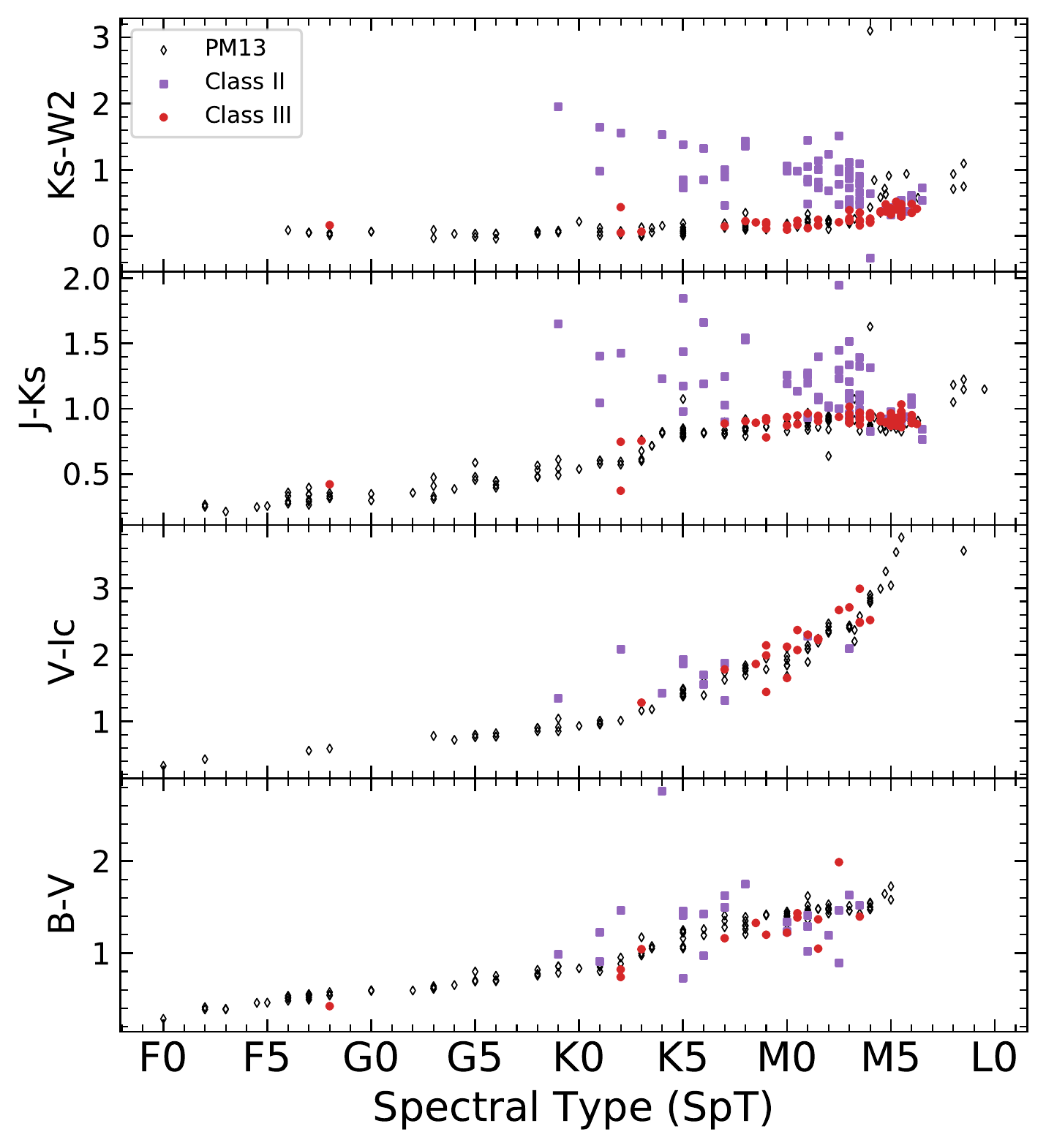}
\caption{Comparison of the \citet{2013ApJS..208....9P} young sample with our members in colors and spectral type. While the Class III sources have a similar behavior to the PM13 sample, the Class II sources both show a red excess in the longer wavelength bands and a blue excess in the shorter wavelength colors.} \label{fig:Figure_5}
\end{figure}

Out of 357 possible members, this designation leads to 131 Class II and 226 Class III stars.
Class III sources are well-suited for our SED method because all the colors can be described by a stellar photosphere model. Class II sources, in the absence of modelling of the IR excess, are better suited for a single color $A_V$ inference method. This is supported by Fig. \ref{fig:Figure_5}, which suggests that Class III sources are similar to the \citet{2013ApJS..208....9P} sample, but the Class II sources show an excess of red in the longer wavelength colors and blue in the shorter wavelength colors. The tight sequence of Class III stars on the color-spectral type relation suggests that there are nearly no Class II stars erroneously flagged with this criterion.

\begin{figure*}
\centering
\includegraphics[trim={0cm 0cm 0cm 0cm},clip,width=\textwidth]{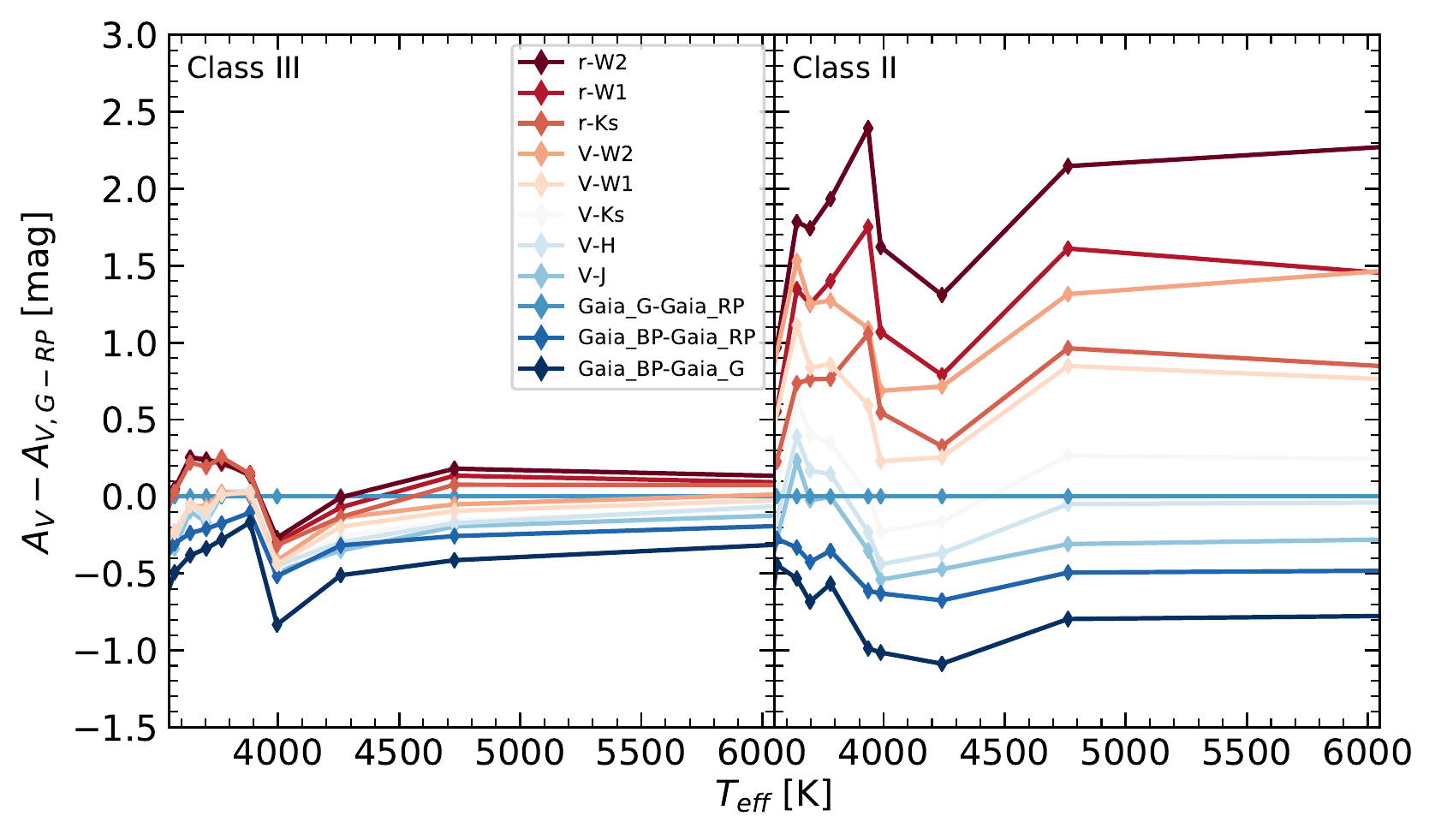}
\caption{A comparison of the extinction estimates of Class III sources and Class II sources, relative to a single-color {\it Gaia} G-RP solution, binned across temperature. Left: the dispersion for Class III sources in our sample increases with temperature. Right: the systematics between colors for Class II sources is very large, spanning magnitudes.} \label{fig:Figure_6}
\end{figure*}

\subsection{Color Tables and Extinction Law} \label{sec:colortables}
Our $A_V$ estimates require models for the photospheric colors of the star and of the physical characteristics of the intervening dust---a set of intrinsic color tables and an extinction law. We first assume the \citet{2013ApJS..208....9P} (PM13) empirical $< 30$ Myr young star table for spectral types F0---M5, with a smooth transition onto the PM13 empirical dwarf color table for stars of spectral type O9---F0 and M5---M9. These tables from PM13 provide the intrinsic colors for only a subset of photometric bands used in our method. To take advantage of the homogenous photometry released by Pan-STARRS and {\it Gaia}, we supplement these intrinsic colors with theoretical color relationships which we can calibrate to an appropriate locus in color space for a self-consistent prescription across multiple colors. We merge this table with the intrinsic colors from \citet{2017AJ....153..188F} to extend our analysis to take advantage of available Pan-STARRS griz data. PARSEC CMD 5 Myr v.3.3 solar metallicity isochrone provides {\it Gaia} colors \citep{2012MNRAS.427..127B} with our Pleiades calibration (See Appendix \ref{sec:gaiacolorsplepra}). We choose a $R_V = 3.1$ extinction law from \citet{1990ARA&A..28...37M}. An extinction law with $R_V = 3.1$ is thought to be most consistent with the nature of dust in the Orion star-forming complex \citep[see Fig.6 in][]{2010ApJ...722.1092D}.

We interpolate linearly over spectral type to obtain effective temperature, color, and bolometric corrections for each star.

\subsection{Extinction Estimation} \label{sec:deredden}

SED fitting and color-based dereddening techniques have often been used to infer stellar parameters extensively in the literature \citep[e.g.,][]{2007ApJ...655..233A,bayo2008,2018AJ....155..196R}. Dereddening techniques are required in the Orion star-forming complex and similar star forming associations in order to estimate the luminosities and ages of these clusters and associations where significant foreground extinction is expected. While SED methods are known to be accurate and precise in open clusters and in joint fits involving multiple stars, it is more challenging to apply these techniques to a cluster as young as {\lamOri}, and to attempt to report extinction solutions for individual stars.

Attempts to perform SED-based dereddening without external temperature information demonstrate a strong degeneracy between extinction and temperature in fits, which is understood to be mitigated by priors or other constraints in work such as \citet{2011MNRAS.411..435B}. In our case, we assume external temperature information, and that the stellar photospheres of our objects are adequately described by our intrinsic young star tables (see Sec. \ref{sec:colortables}). When there are no significant systematics in any of the constituent bands, and the template SED measures the photosphere of the star at a chosen $T_{\mathrm{eff}}$, the increased number of data points reduces the uncertainty of $A_V$. This is because when the intrinsic stellar template is provided, the SED fitting technique reduces to a $\chi ^2$ minimization problem across expected and observed colors.

With a representative template for the colors of a star at a fixed temperature, each color measurement provides a color excess relative to the intrinsic color of the star---therefore, a measurement of $A _\lambda$. With an extinction law, each color provides a covariant measurement of $A_V$.
From the extinction law, we define the extinction vector:
\begin{equation}
\vect{\hat{A}}_{i-j} = \frac{A_i}{A_V} - \frac{A_j}{A_V} ,
\end{equation}
where $i$ and $j$ correspond to different photometric bands, and $\frac{A_i}{A_V}$ to the extinction law evaluated at band $i$. We also define the color excess vector:
\begin{equation}
\vect{E}_{i-j} = (i-j) - (i-j)_0 .
\end{equation}
Where $(i-j)$ corresponds to the observed color, and $(i-j)_0$ the intrinsic color for a star with a reported spectral type. With these definitions, we denote a color residual vector:
\begin{equation}
\vect{r} = A_V * \vect{\hat{A}} - \vect{E}. \label{eqn:resids}
\end{equation}
Then the $\chi ^2$ statistic is defined,
\begin{equation}
\chi _{SED} ^2 \equiv \vect{r}^T \matr{C}^{-1} \vect{r} , \label{eqn:chisq_mat}
\end{equation}
where $\vect{r}$ is the vector of color residuals and $\matr{C}$ is the covariance matrix between pairs of colors, whose derivation is given in Appendix~\ref{sec:errorprop}.

In Appendix \ref{sec:plepra}, particularly Fig. \ref{fig:Figure_B3}, a demonstration of the technique can be seen applied to the Pleiades and Praesepe. As the $T_{\mathrm{eff}}$'s are calculated from colors involving V and J, H, Ks in this example, the color vector chosen is not exactly the same as in our analysis (Sec. \ref{sec:colorsed}). Nevertheless, the use of multiple colors in our SED fit drives down the dispersion of extinctions, dramatically in Praesepe, and more modestly in the Pleiades.

\subsection{SED Fitting} \label{sec:sedfitting}
\subsubsection{Definitions and Methodology} \label{sec:colorsed}
For Class III sources, we can define an estimator for the weighted mean of $A_V$ across all colors for a single star from Eqns \ref{eqn:resids} \& \ref{eqn:chisq_mat}. We first find the minimum of the $\chi^2$ statistic, noting for our definitions of the column vectors $\vect{\hat{A}}$ and $\vect{\hat{E}}$, that $\vect{\hat{A}}^T \matr{C}^{-1} \vect{E} = \vect{E}^T \matr{C}^{-1} \vect{\hat{A}}$:
\begin{equation}
\frac{\partial \chi _{SED} ^2}{\partial A_V} = - 2 \vect{\hat{A}}^T \matr{C}^{-1} \vect{E} + 2 A_V * \vect{\hat{A}}^T \matr{C}^{-1} \vect{\hat{A}} ,
\end{equation}
Setting this to zero leads to the the estimator:
\begin{equation}
\left< A_V \right> _{\mathrm{SED}} \equiv \frac{ \vect{\hat{A}}^T \matr{C}^{-1} \vect{E} }{ \vect{\hat{A}}^T \matr{C}^{-1} \vect{\hat{A}} }.
\end{equation}
We verify that this is a minimum in $\chi^2$ by taking another derivative with respect to $A_V$:
\begin{equation}
\frac{\partial ^2 \chi _{SED} ^2}{\partial A_V ^2} = 2 \vect{\hat{A}}^T \matr{C}^{-1} \vect{\hat{A}} .
\end{equation}
This value is always positive when the covariance matrix is diagonal (when no bands are repeated in a color system). However, for nontrivial color combinations, we check this value to ensure that we are choosing the $\chi ^2$ minimum in a well-defined color system (Appendix \ref{sec:errorprop}).

We restrict colors for the $A_V$ SED calculation to a self-consistent set using our calibration in the Pleiades \& Praesepe (Appendix \ref{sec:plepra}). These colors are the {\it Gaia} colors (BP-G, BP-RP, and G-RP), colors involving Pan-STARRS photometry (r-J, r-H, r-Ks, r-W1, r-W2, i-J, i-H, i-Ks, i-W1, i-W2), and colors involving Johnson-Cousins photometry (V-J, V-H, V-Ks, V-W1, V-W2).

For Class II sources, we simplify this treatment by using the single color {\it Gaia} G-RP. Out of the colors to choose for a single color extinction model, we use {\it Gaia} G-RP because it is a broadband color available for our Class II sources which mitigates the non-extinction systematics that blue or NIR bands might be sensitive to (such as accretion, activity, or disks). We choose G-RP to be the single color we use for Class II sources, and benchmark the performance of other colors against it. In Fig. \ref{fig:Figure_6}, we plot a series of binned single-color extinctions for our cluster members and subtract off the extinction inferred from the {\it Gaia} G-RP color. Since a large number of these stars have disks or accretion signatures, the underlying template is no longer a good fit for the photosphere in some of the red and blue colors; since the $A_V$ difference increases at redder colors, the extinction estimates are no longer reliable as they can be biased by the circumstellar disk.

\begin{figure}
\centering
\includegraphics[trim={0cm 0cm 0cm 0cm},clip,width=\columnwidth]{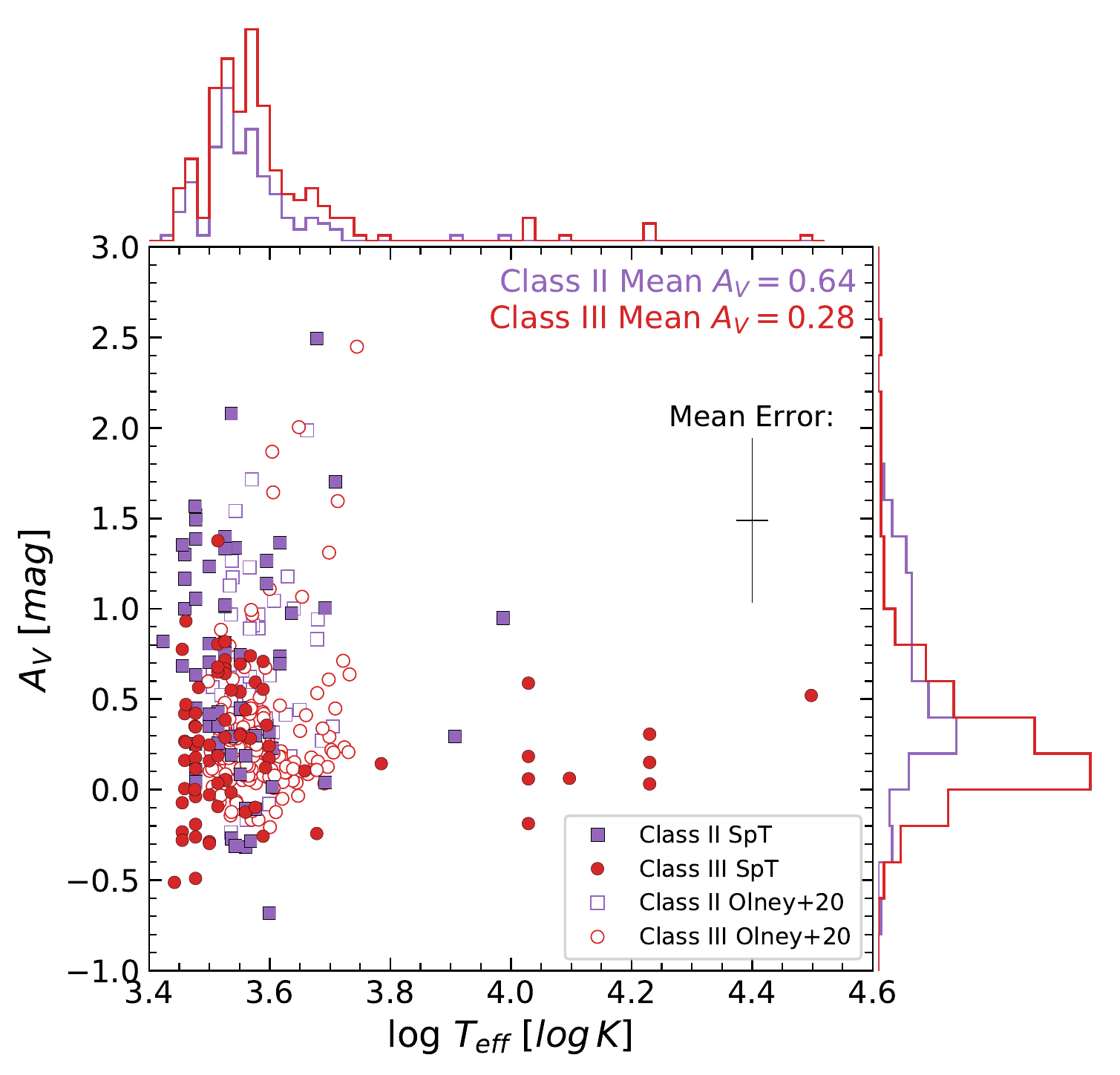}
\caption{$T_{\rm eff}$---$A_V$ relationship for our population in {\lamori}, color-coded by Class diagnostic. Class II members have a much larger dispersion and have a systematically higher extinction value than the Class III stars. The variation of $A_V$ is relatively flat with $T_{\mathrm{eff}}$, suggesting consistency in our treatment for estimating star-by-star $A_V$.} \label{fig:Figure_7}
\end{figure}

In Fig. \ref{fig:Figure_7}, we show the distribution of inferred extinctions as a function of effective temperature. For the \citet{2020AJ....159..182O} sample, it is interesting to note that we infer systematically lower extinctions for the Class III sources than the Class II sources. Note that we infer a range of negative extinctions as well as positive extinctions. Negative extinctions are unphysical, but they can be produced from random photometric errors or inconsistencies between spectral types, the photometry, and the extinction model. Since the negative extinctions considered in this way are close to 0 ($< 1 \sigma$), we believe that in most cases they simply reflect random measurement uncertainties. We therefore consider these values and errors to probe the underlying age distribution, without truncating these extinctions at 0 mag. This has the practical effect of applying a de-bluening for these few stars, but avoids artificially biasing the $A_V$ distribution or modifying the resultant age distribution.

Negative extinction solutions in these stars may have to do with the physical effects of spots or activity.
These physical effects are a likely explanation given the youth of this cluster, with one quarter to half of members showing signs of disks, spots, and active accretion \citep[e.g.][]{2015AJ....150..100K,2012A&A...547A..80B,2011A&A...536A..63B,2008A&A...488..167S}. Alternate explanations for the negative extinction solutions may include underestimated systematic errors in spectral types or mismatches between stellar photospheres and intrinsic colors. We explore the possibility of a systematic for all our stars in Sec. \ref{sec:systematics}, but note that a mismatch between our colors is unlikely since our population appears to follow the empirical trends of PM13 in Fig. \ref{fig:Figure_5}, and the {\it Gaia} colors are calibrated in Appendix \ref{sec:plepra}.

\subsubsection{Error Estimation} \label{sec:errorestimation}

We propagate errors within our code to estimate the individual stellar age and mass errors within our model, and infer cluster values for the age uncertainties and age spread. Mechanically, we obtain a $T_{\mathrm{eff}}$ error and a corresponding luminosity error for each star, and we obtain this primarily by doing linear error propagation.

First, in order to estimate $A_V$ error, we assume that the propagated spectral type errors and photometric errors are independent from each other.
We derive an $A_V$ error estimate using the spectral type and photometric errors, reporting the extinction error as the standard error of the mean using our color system:
\begin{equation}\label{eqn:sigAv}
\sigma _{A_V} = \left( \left[\left( \frac{ \vect{\hat{A}}^T \matr{C}^{-1} \vect{N} }{ \vect{\hat{A}}^T \matr{C}^{-1} \vect{\hat{A}} } \right)^2 \sigma _{\textnormal{SpT}} ^2 \right]^{-1} + \left[\vect{V}^T \, \matr{C} \, \vect{V} \right] ^{-1} \right) ^{- \frac{1}{2}},
\end{equation}
where $\vect{N}_{i-j} = \frac{\partial \,  (i-j)_{0}}{\partial \, \textnormal{SpT}}$ is the vector of partial derivatives of the intrinsic color table with respect to spectral type, $\vect{V}_{i-j} = \frac{\partial \,  A_V}{\partial \, i-j} = \left( \frac{A_i}{A_V} - \frac{A_j}{A_V} \right)^{-1}$ is the vector of partial derivatives of $A_V$ with respect to each color, and $\matr{C}$ the corresponding covariance matrix (Appendix \ref{sec:errorprop}).
We estimate $\vect{N}$ through interpolations across the empirical color table.
\begin{figure}
\centering
\includegraphics[trim={0cm 0cm 0cm 0cm},clip,width=\columnwidth]{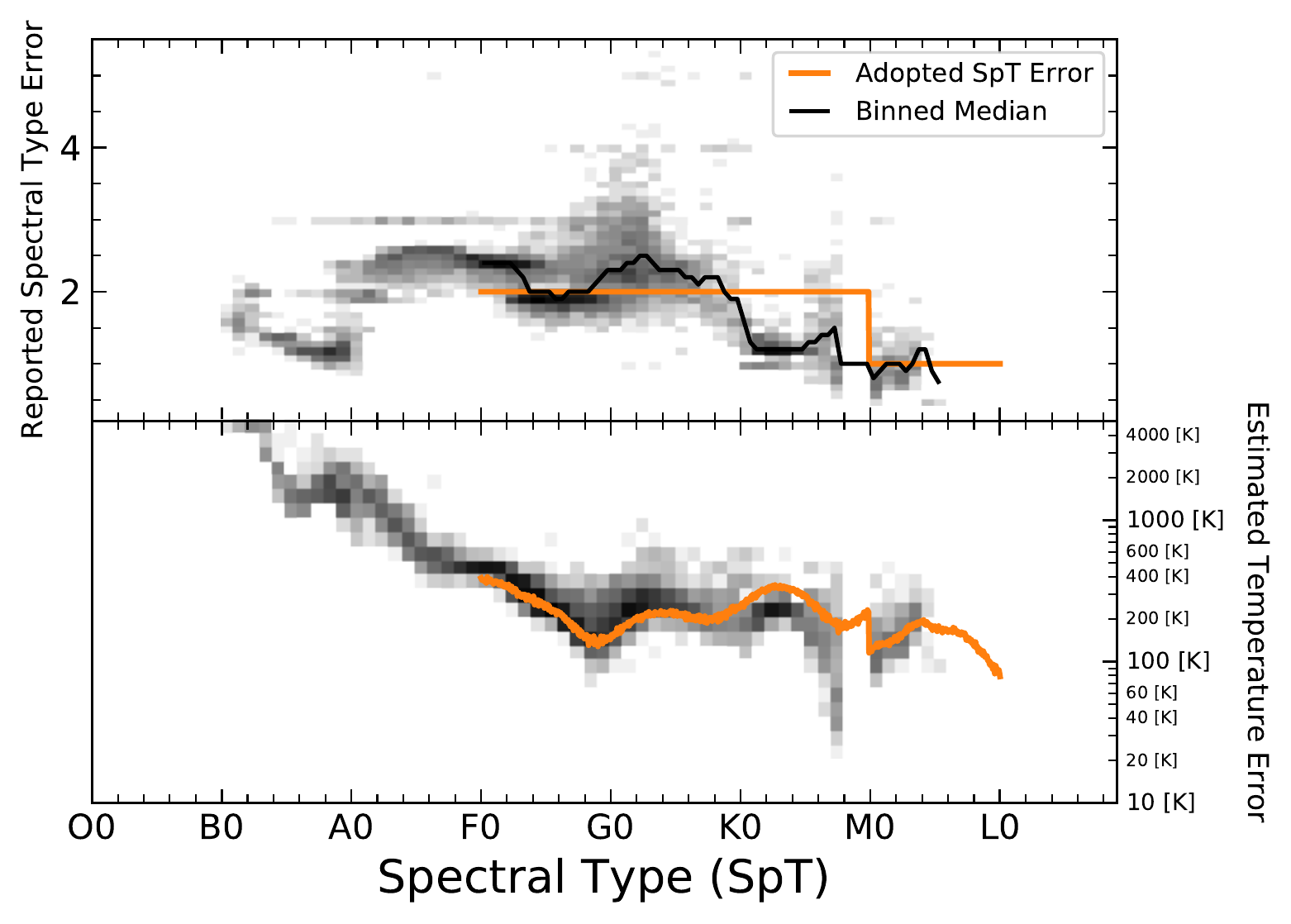}
\caption{Top: Comparison between literature spectral type uncertainties reported by YSOC and our proposed ``default'' spectral type error law for any data missing a reported spectral type error. Bottom: Translation of spectral type error to temperature uncertainty. In most cases our adopted default spectral type errors are close to the median value, and in the effective temperature plane our errors hover around $100$---$300$ K. The shading is according to logarithmic bin density.} \label{fig:Figure_8}
\end{figure}

For stars where $\sigma _{\textnormal{SpT}}$ is not provided, we assume ``default'' errors of 2 subclasses for stars from O-M0, and 1 subclass for stars later than M0. This corresponds to a temperature error of roughly $2000$ K ($<$ B9), $400$ K (A0---A9), $200$ K (F0---G9), $300$ K (K0---K5), $150$ K (M0---M5), and $100$ K (M5---M9). In Fig. \ref{fig:Figure_8} we check these errors against reported literature spectral type errors for similar young stars in the YSOC database \citep{hillenbrand2021csss}.

For stars in the database where $\sigma _{\textnormal{phot}_i}$, $\sigma _{\textnormal{plx}}$ are not provided, we assume ``default'' errors in photometry and parallax:
\begin{equation}
\begin{cases}
\sigma_{\textnormal{default phot}} &= 0.01 \; \text{[mag] for all bands}, \\
\sigma_{\textnormal{default log dist [pc]}} &= 0.1 \; \text{[dex] for all distances}.
\end{cases}
\end{equation}
Since the bulk of stars in our sample have well-defined photometric errors and {\it Gaia} parallaxes, these default errors are used for a very small number of sources.

\subsubsection{Comparison with YSOC Literature $A_V$} \label{sec:ysoc_lit}

\begin{figure}
\centering
\includegraphics[trim={0cm 0cm 0cm 0cm},clip,width=\columnwidth]{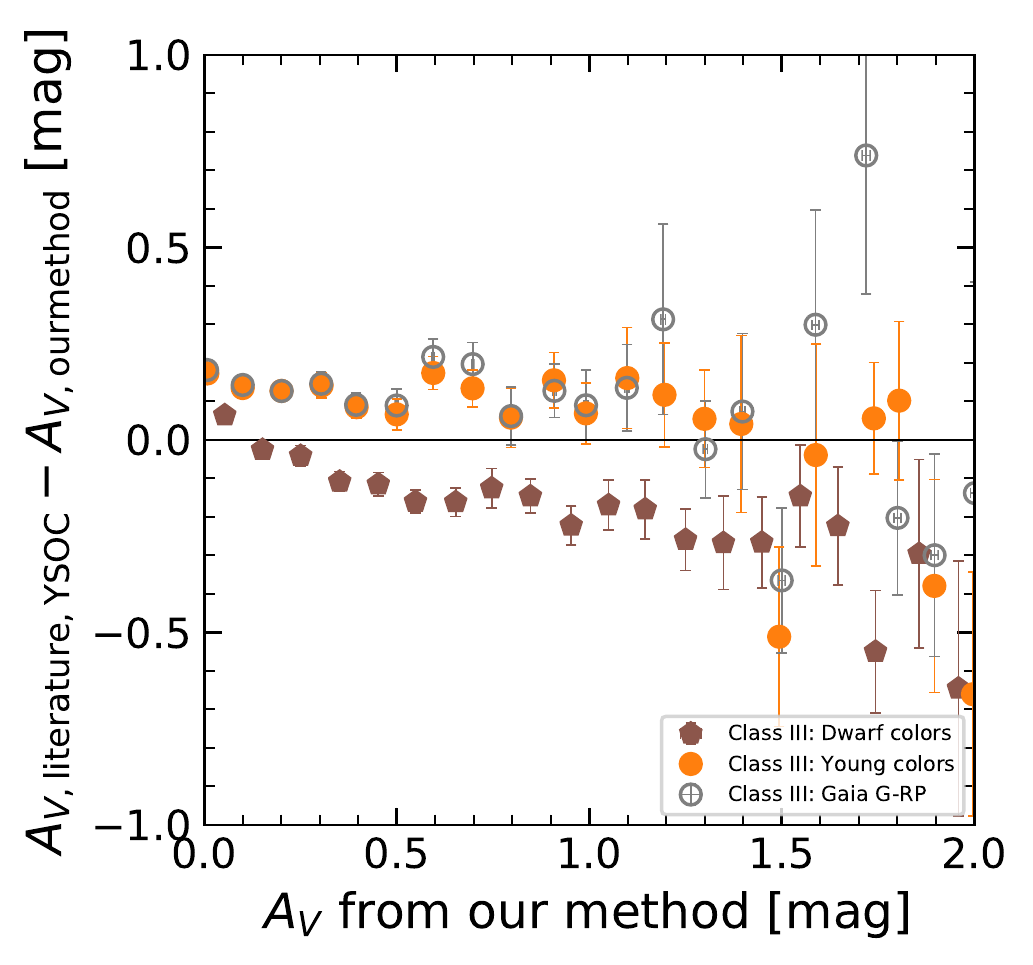}
\caption{Comparison between methods used in this paper and aggregate literature sources in the YSOC database for Class III sources. Symbols are binned across our SED method's extinction, with a standard error represented by the bar and the dot landing on the bin mean.} \label{fig:Figure_9}
\end{figure}

In Fig. \ref{fig:Figure_9} we compare our extinction estimates across all available clusters in the YSOC database to available literature values for a variety of our pipeline methods \citep{hillenbrand2021csss}. Though there may be zero-point offsets in extinctions calculated with different methods and different intrinsic stellar templates, the young (orange) and single-color (gray unfilled) are consistent with each other. The usage of dwarf colors in extinction inference has a much larger systematic effect on $A_V$.

\subsection{Inferring $L$, $T_{\rm eff}$} \label{sec:put_hrd}

After an $A_V$ is derived for a star, we determine luminosity from the V-band photometry, bolometric correction, parallax, and extinction. For stars which do not have V-band photometry we construct a spectrum that matches the existing dereddened SED using our color table and infer a synthetic V-band magnitude.
\begin{equation}
log \, L/L_{\odot} = \frac{2}{5} \left[ M_{\mathrm{bol}, \odot} - \left(  \mathrm{BCV} +  M_{V \mathrm{, dered}} \right) \right] ,
\end{equation}
the uncertainties on the observables lead to:
\begin{equation}
\sigma _{log \, L/L_{\odot}} = \sqrt{\sigma_{A_V}^2 + \sigma_{V}^2 + \sigma_{\mathrm{BCV}}^2 + \left({2*\sigma_{d}/d}\right)^2}.
\end{equation}
Where $\sigma_{V}$ corresponds to the quadrature-summed random errors from the V-band photometry (or the synthetic V-band photometry from a match to the SED at the closest band). Since our SED fit does not explicitly consider departures from our adopted color table, all stars of the same temperature are assigned the same BCV; the mismatch between stellar BCV's represents a systematic, but we do not try to investigate it in this work.
$T_{\rm eff}$ for the spectral type sources in {\lamori} is computed by interpolating on the young stellar SpT---$T_{\rm eff}$ relation from \citet{2013ApJS..208....9P}. Errors in $T_{\mathrm{eff}}$ for these stars are found by propagating the uncertainty in spectral type in the interpolated relation.

For the APOGEE Net stars, we infer $T_{\mathrm{eff}}$ by applying the temperature correction from Sec. \ref{sec:temperature}. For these stars we inflate temperature errors reported in APOGEE Net by a factor of five to make them comparable with temperature errors inferred with spectral types, consistent with published advice \citep{2020AJ....159..182O}.

\begin{figure*}
\centering
\includegraphics[trim={0cm 0cm 0cm 0cm},clip,width=0.8\textwidth]{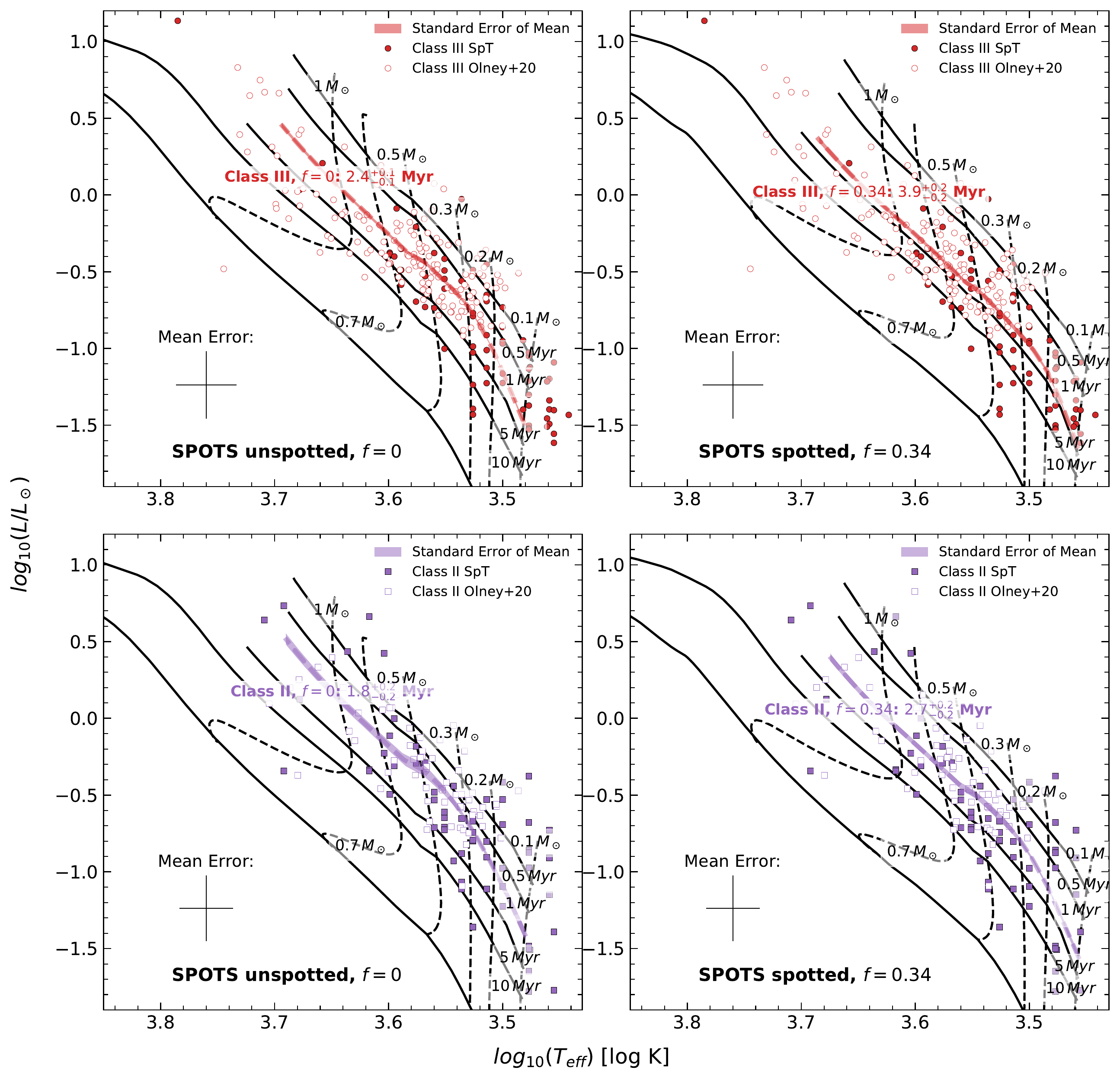}
\caption{HR Diagram of {\lamOri} members with bands from the fiducial results from Table \ref{tab:ages}, clipped to the mass range (0.1---1.3 $M_\odot$) of the SPOTS isochrones. Left to right: SPOTS unspotted and spotted, 50\% covering fraction models. Top to bottom: Class III \& II sources. Literature spectral type sources are filled symbols while APOGEE Net sources are unfilled symbols. Class II sources are plotted in purple and Class III sources are plotted in red. Translucent bands with a dashed line running through them represent the appropriate SPOTS inferred age for each Class. Spotted models predict consistent ages for the lowest mass stars, which in the unspotted case appears erroneously younger than the rest of the cluster.} \label{fig:Figure_10}
\end{figure*}

\begin{figure*}
\centering
\includegraphics[trim={0cm 0cm 0cm 0cm},clip,width=0.8\textwidth]{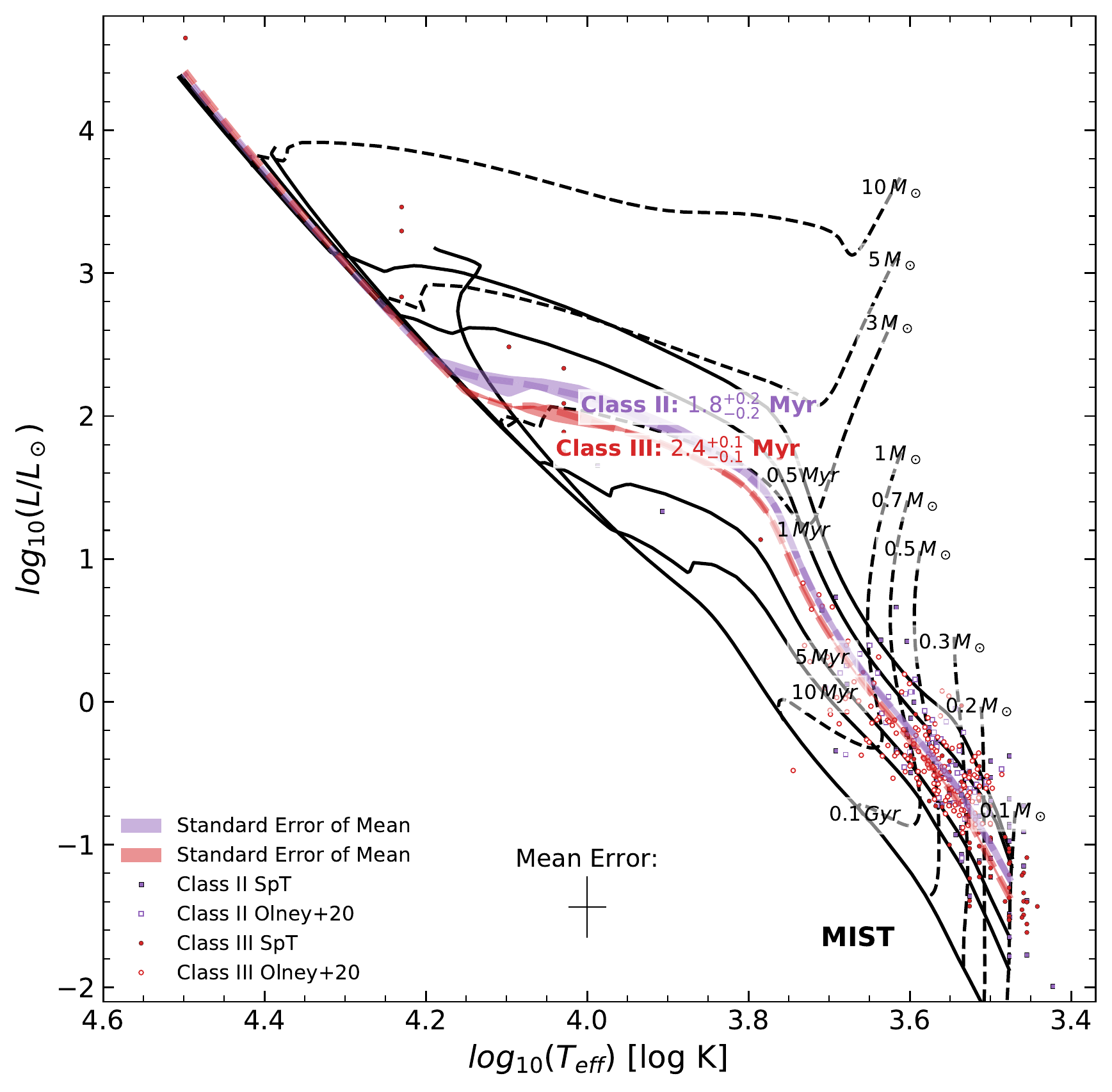}
\caption{Full HR Diagram of {\lamOri} members with bands from the results run on MIST from Table \ref{tab:ages}. Purple squares correspond to Class II sources, while red circles correspond to Class III sources. Filled symbols correspond to stars from our literature sample, while unfilled symbols correspond to stars from APOGEE Net.} \label{fig:Figure_11}
\end{figure*}

\begin{figure}
\centering
\includegraphics[trim={0cm 0cm 0cm 0cm},clip,width=\columnwidth]{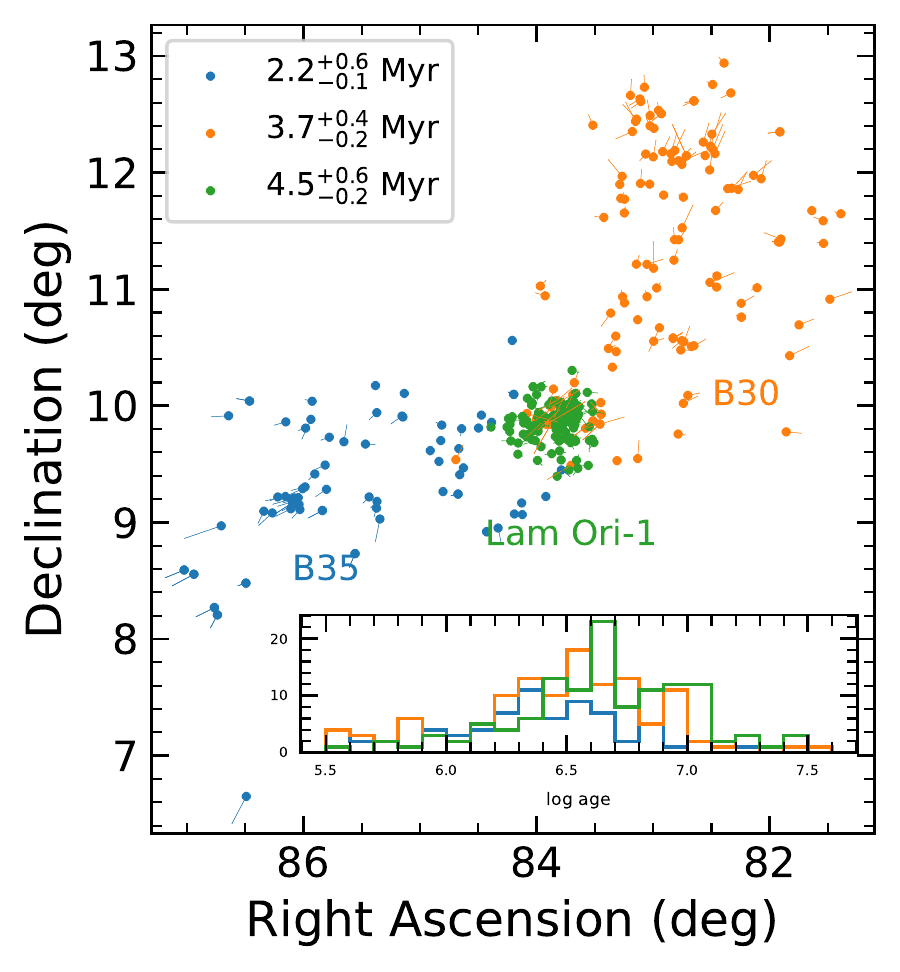}
\caption{Kinematic regions associated with {\lamOri} and assorted proper motion vectors and age distributions, with the standard error of the mean provided as error bounds. While the distributions have different means, with B30 and B35 noticeably younger, the dispersion in luminosity and age is similar between {\lamOri} and its subregions as well. The KS statistic suggests that B35 and B30 are indistinguishable at the $5\%$ level, but are easily distinguished from the center.} \label{fig:Figure_12}
\end{figure}

\subsection{Inferring Masses and Ages} \label{sec:massesages}

We construct a Bayesian framework to provide an estimate on mass and age with input luminosities and effective temperatures for individual stars.
This analysis is on the theoretical plane with luminosity and effective temperature because these stellar properties are an output of our SED-fitting method. This allows us to test the underlying luminosity and temperature predictions from a variety of stellar models.

We derive individual masses for all the members of the {\lamori} cluster by using a variety of models, but choose as our fiducial model SPOTS (0.1---1.3 $M_\odot$) with $f_{\mathrm{spot}} = 0.34$ \citep{2020ApJ...891...29S,2015ApJ...807..174S}. We make the starspot fraction choice \textit{a priori} because these stars are likely to be heavily spotted. Fits from young eclipsing binary systems and cluster members in Upper Sco suggest that $f_{\mathrm{spot}} \sim 0.3$---$0.5$ effectively reproduces the slope of the mass---age relationship for cluster members and for individual EB's \citep{2020ApJ...891...29S,2019ApJ...872..161D,2017ApJ...836..200G}. Our application of the $f_{\mathrm{spot}}=0.34$ and $f_{\mathrm{spot}}=0.5$ isochrones were more consistent in Upper Sco across low and high mass bins, and the $f_{\mathrm{spot}}=0.34$ spotted isochrones were better fits to eclipsing binary data in that cluster \citep{2020ApJ...891...29S,2019ApJ...872..161D}.

In Fig. \ref{fig:Figure_10} we demonstrate the effects of these starspot models for our Class II \& III stars. The Class II population appears younger than the Class III population with the lowest mass members better accounted for by the spotted isochrones. The curvature of the unspotted isochrones below $0.1 M_\odot$ provide ages that are systematically younger than the population of older stars (top \& bottom left). The age consistency across stars of different masses is in good agreement with prior results \citep{2020ApJ...891...29S,2019ApJ...884...42S,2019ApJ...872..161D}. In Fig. \ref{fig:Figure_11} we plot our members across the entire mass range (B0---M9) on a set of MIST pre-MS isochrones. We choose to plot the temperature and luminosity values from our table against the MIST isochrones because they go to much higher mass than the SPOTS isochrones, which are limited to 0.1---1.3 $M_\odot$---this expanded range in mass makes it possible to see the results of the dereddening procedure for the whole sample in {\lamori}. The resulting band of ages largely spans the range from 1---5 Myr, a much tighter band than individual dereddening procedures in {\lamori} members with ages ranging from 5---40 Myr (e.g. \citet{2011A&A...536A..63B}).

\subsubsection{Priors}

We use a mass prior with a \citet{2003PASP..115..763C} IMF:

\small
\begin{equation}
\frac{d \, n}{d \, log \, M} = \begin{cases}
    0.093 \times \mathrm{exp} \left[ - \frac{\left( log \, \frac{M}{M_\odot} - log \, 0.2 \right) ^2}{2 * 0.55 ^2} \right]& \text{if $\frac{M}{M_\odot} < 1$},\\
    0.041* \left( M/M_\odot \right)^{-1.35} & \text{otherwise}.
  \end{cases}
\end{equation}
\normalsize

Since the age distribution is unknown to us, it suffices to choose a uniform age prior with a maximum age:

\begin{equation}
\frac{d \, n}{d \, log \, A} = \begin{cases}
    1, & \text{if $A < 100 \; \mathrm{Myr}$},\\
    0 & \text{otherwise}.
  \end{cases}
\end{equation}

The choice of these priors does not significantly perturb the age and mass estimation; in tests, the choice of a uniform mass prior has less than a $\sim$0.01 dex effect on the inferred ages, and an effect of $\sim$0.01 $M_\odot$ on the inferred masses.

\begin{deluxetable*}{ccc}
\tabletypesize{\small}
\tablenum{2}
\tablecaption{Inferred Cluster Age for the {\lamori} cluster across models\label{tab:ages}}
\tablewidth{0pt}
\tablehead{
\colhead{Model} &
\colhead{Class II Age (Myr)} &
\colhead{Class III Age (Myr)}
}
\startdata
SPOTS $f=0$ & $1.8 ^{+0.2} _{-0.2} \, {\mathrm{ Myr}}$ & $2.4 ^{+0.1} _{-0.1} \, {\mathrm{ Myr}}$ \\
\fbox{SPOTS $f=0.34$} & $2.7 ^{+0.2} _{-0.2} \, {\mathrm{ Myr}}$ & $\boxed{3.9 ^{+0.2} _{-0.2} \, {\mathrm{ Myr}}}$ \\
SPOTS $f=0.5$ & $3.6 ^{+0.3} _{-0.3} \, {\mathrm{ Myr}}$ & $5.0 ^{+0.3} _{-0.3} \, {\mathrm{ Myr}}$ \\
MIST (standard) & $1.8 ^{+0.1} _{-0.1} \, {\mathrm{ Myr}}$ & $2.4 ^{+0.1} _{-0.1} \, {\mathrm{ Myr}}$ \\
BHAC15 (standard) & $1.7 ^{+0.1} _{-0.1} \, {\mathrm{ Myr}}$ & $2.5 ^{+0.1} _{-0.1} \, {\mathrm{ Myr}}$ \\
DSEP (standard) & $1.7 ^{+0.1} _{-0.1} \, {\mathrm{ Myr}}$ & $2.4 ^{+0.1} _{-0.1} \, {\mathrm{ Myr}}$ \\
Feiden DSEP (magnetic) & $3.3 ^{+0.2} _{-0.2} \, {\mathrm{ Myr}}$ & $4.5 ^{+0.2} _{-0.2} \, {\mathrm{ Myr}}$ \\
PARSEC (empirical Dwarf) & $2.4 ^{+0.2} _{-0.2} \, {\mathrm{ Myr}}$ & $3.2 ^{+0.1} _{-0.1} \, {\mathrm{ Myr}}$ \\
\enddata
\tablecomments{Inferred cluster ages, along with the standard errors of the mean. Classic stellar isochrones for Class III stars predict ages of 2-3 Myr, while spotted and magnetic isochrones produce systematically older estimates ranging from 4-5 Myr. Our fiducial spotted model, SPOTS with a spot filling fraction of 34\%, is shown here boxed.}
\end{deluxetable*}

\subsubsection{Obtaining Posteriors}

In order to define a probability distribution and a posterior on our age-mass grid, we define a statistic on our theoretical HRD space:
\begin{equation}
\chi ^2 _{T_{\mathrm{grid}}, L_{\mathrm{grid}}} = \frac{\left( log \, T/T_{\mathrm{grid}} \right)^2}{\sigma _{log \, T} ^2} + \frac{\left( log \, L/L_{\mathrm{grid}} \right)^2}{\sigma _{log \, L / L_\odot } ^2}.
\end{equation}
We define a grid uniformly sampled on $log \, M$, $log \, A$, such that $\Delta \, log \, M$ and $\Delta \, log \, A$ are independent of grid point. We perform a 2D interpolation over mass and age to get $T(M, A)$ and $L(M, A)$ to evaluate this posterior,
\begin{eqnarray}
P \left(M_{\mathrm{grid}},A_{\mathrm{grid}} \right) = P \left(M_{\mathrm{grid}} \right) \, P \left(A_{\mathrm{grid}} \right) \times \nonumber \\
\frac{\mathrm{exp} \, \left( - \frac{1}{2} \chi ^2 _{T(M_{\mathrm{grid}},A_{\mathrm{grid}}), L(M_{\mathrm{grid}},A_{\mathrm{grid}})} \right)}{\sum _{M',A'} ^{grid} P(M'_{\mathrm{grid}}, A'_{\mathrm{grid}}) }.
\end{eqnarray}
We then marginalize over the distributions:
\begin{equation}
P \left( M_{\mathrm{grid}} \right) = \sum _{A_{\mathrm{grid}}} P \left( M_{\mathrm{grid}}, A_{\mathrm{grid}} \right) \Delta \, log \, A,
\end{equation}
\begin{equation}
P \left( A_{\mathrm{grid}} \right) = \sum _{M_{\mathrm{grid}}} P \left( M_{\mathrm{grid}}, A_{\mathrm{grid}} \right) \Delta \, log \, M.
\end{equation}
To construct $1 \sigma$ confidence intervals, we then pick the smallest parameter range in the marginal cumulative probability distribution function that includes $68 \%$ of the probability.

Because our extinctions are set by our choice of color table and our external spectral types, we can obtain individual stellar masses and ages with associated uncertainties. We report individual ages, masses, luminosities, and temperatures for our {\lamori} cluster members in Table \ref{tab:stellarparams}.

\section{Results} \label{sec:results}

\subsection{Cluster Age} \label{sec:clusterage}

Since we produce individual ages with unique uncertainties for each member in our sample, we are able to do statistics on the underlying distribution of stellar ages to determine a best-fit age for the cluster for a variety of different underlying theoretical models. We include the following stellar isochrones: SPOTS \citep{2020ApJ...891...29S}, PARSEC \citep{2015MNRAS.452.1068C}, BHAC15 \citep{2015A&A...577A..42B}, DSEP with pre-MS physics \citep{2008ApJS..178...89D}, Feiden magnetic DSEP \citep{2014ApJ...789...53F,2012ApJ...761...30F}, and MIST \citep{2016ApJ...823..102C}.

\begin{figure*}
\centering
\includegraphics[trim={0cm 0cm 0cm 0cm},clip,width=\textwidth]{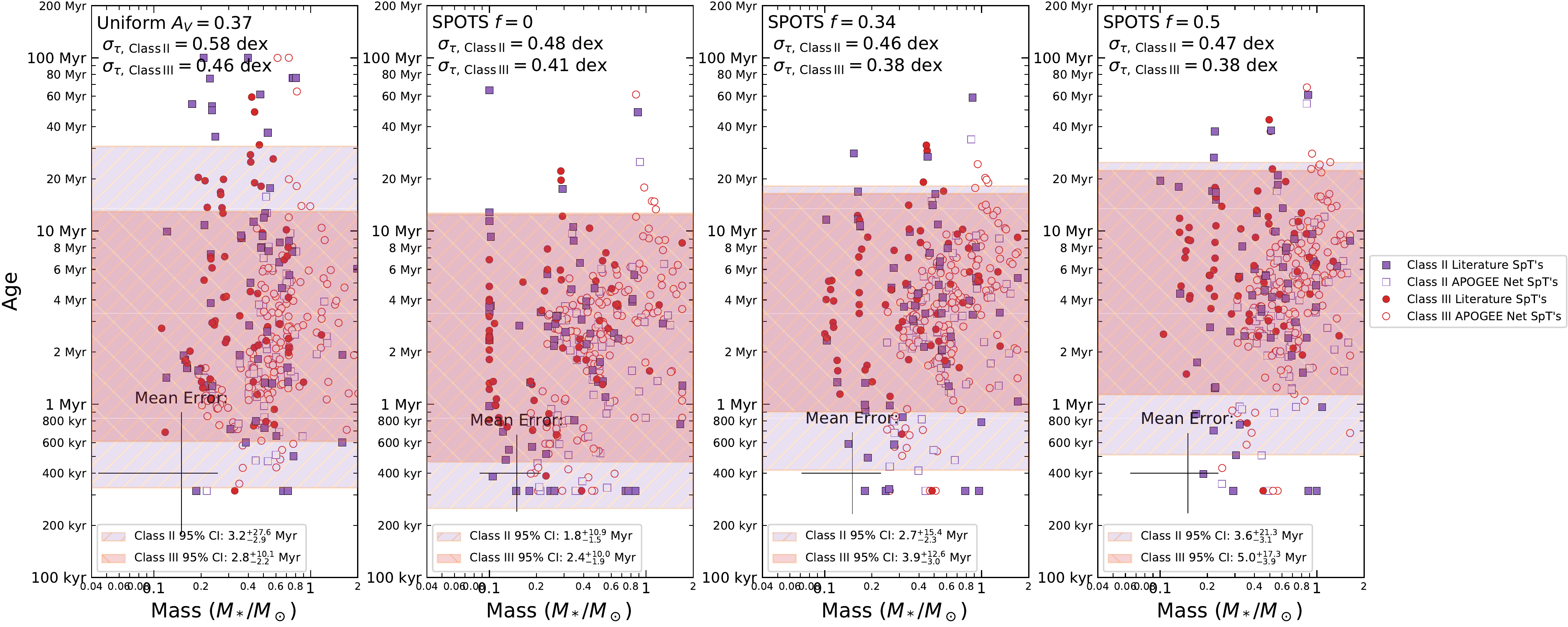}
\caption{Age-mass estimates for {\lamori}. A comparison between the ages derived using a single $A_V = 0.37$ mag and from our SED technique, with bands indicating the 95\% confidence interval for each population. Using a single extinction value for the whole cluster significantly increases the age spread in the Class II sources. Filled symbols correspond to SpT's derived from the literature sample, while unfilled symbols correspond to SpT's from APOGEE Net. \label{fig:Figure_13}}
\end{figure*}

{\lamOri} is well known to be a cluster with kinematically distinct populations, including the regions in B30 and B35, which are traditionally considered the ``cloud'', and the center \citep{dolan2001, 2018AJ....156...84K}. In Fig. \ref{fig:Figure_12}, we find that the B35 and B30 populations do appear to be younger, with mean ages of $2.2$ and $3.7$ Myr, respectively, but the spread in ages within these subregions are still large enough that many stars in the distributions appear to overlap in age. Running a Kolmogorov-Smirnov test on the distributions finds a KS statistic of $2 \times 10^{-3}$ between B30 and the center, $3 \times 10^{-6}$ between B35 and the center, and $6\%$ between B30 and B35. This suggests that the B35 and B30 regions appear to be drawn from a distribution that is distinct from the stars in the center of {\lamori}, but do not appear to be drawn from distributions that are different from each other at a $5\%$ level, similar to the analysis done by \citet{dolan2001}. Since the distributions have largely the same dispersion in log age, the subsequent analysis for the age spread is unlikely to change significantly when considering these regions.

We report in Table \ref{tab:ages} the binned inferred Bayesian ages, the corresponding standard error of the mean, and the age spread for each family of models. We provide solutions for Class II and Class III sources in independent calculations. This is motivated by the different methods we used in Sec. \ref{subsec:prefilter} \& \ref{sec:colorsed}, and the expected difference in quality of our extinction estimates in those groups (Fig. \ref{fig:Figure_6}). We expect our Class III sources to have better constrained extinction uncertainties than our Class II sources for the reasons discussed in Sec. \ref{subsec:prefilter}.
As a result, we believe that the age estimate derived with the spotted isochrones on the Class III sources, $3.9 ^{+0.2} _{-0.2}$ Myr, is the strongest individual diagnostic that we have on the age of the cluster.

There are variations in inferred cluster mean age between each of the individual isochrones and between the Class II \& III populations.
The largest discrepancy between inferred ages comes from the shift of spotted isochrones towards older ages. The 34\% spotted (age: $3.9 ^{+0.2} _{-0.2}$ Myr), 51\% spotted isochrones (age: $5.0 ^{+0.3} _{-0.3}$ Myr), and the magnetic DSEP isochrones (age: $4.5 ^{+0.2} _{-0.2}$ Myr) yield significantly older ages. This is because of the strong shift in L,T space of spotted and magnetic isochrones in SPOTS \citep{2020ApJ...891...29S,2015ApJ...807..174S} and magnetic DSEP \citep{2019ApJ...884...42S,2016A&A...593A..99F,2014ApJ...789...53F,2012ApJ...761...30F} isochrones. The shift from standard DSEP to Feiden magnetic DSEP has a similar effect---shifting from $2.4 ^{+0.1} _{-0.1}$ to $4.5 ^{+0.2} _{-0.2}$ Myr in the Class III population. This strong shift suggests that the perturbative effect of magnetism and spots in young clusters may be more significant than variations between different families of evolutionary tracks.

\begin{figure*}
\centering
\includegraphics[trim={0cm 0cm 0cm 0cm},clip,width=\textwidth]{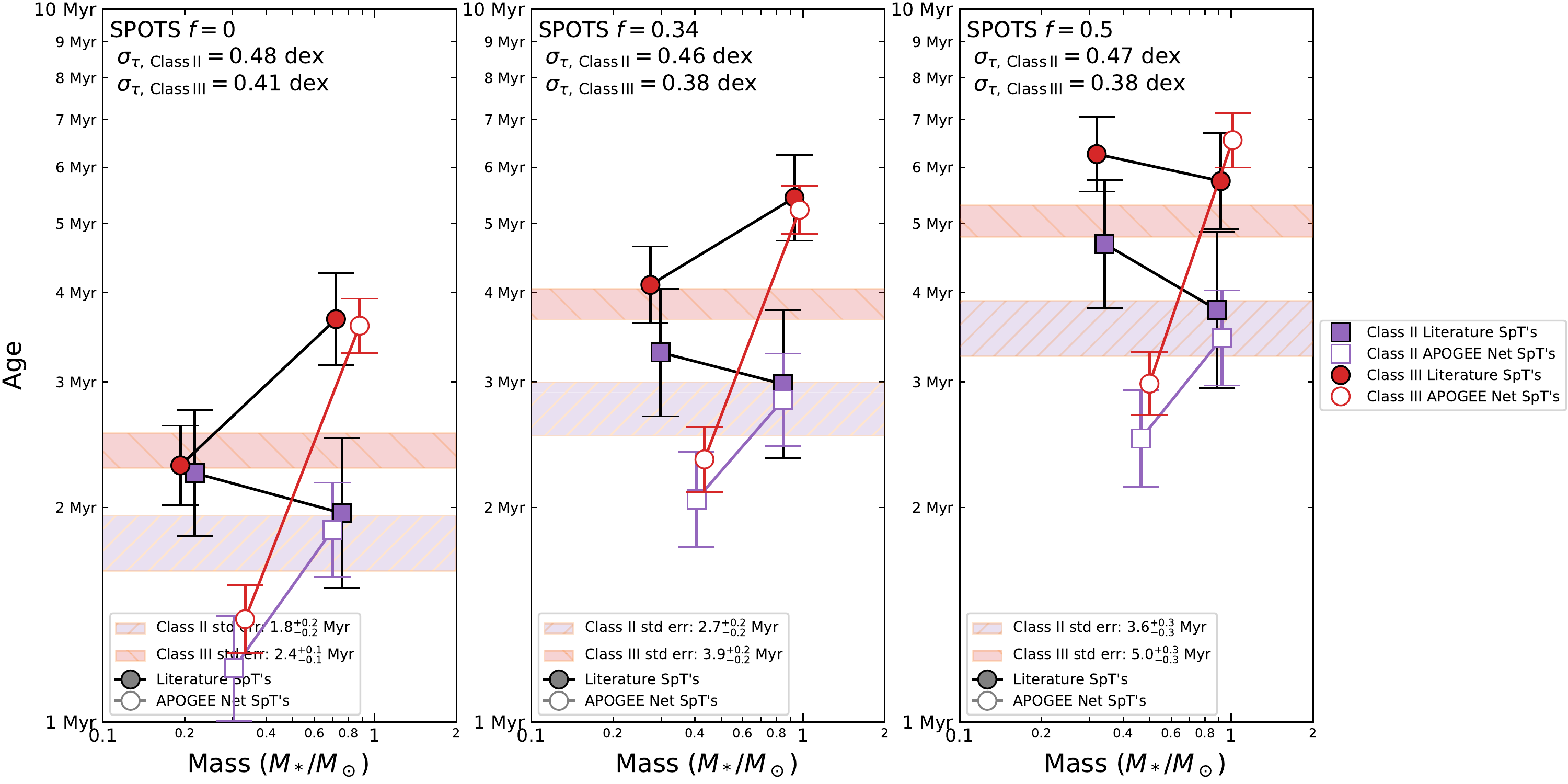}
\includegraphics[trim={0cm 0cm 0cm 0cm},clip,width=\textwidth]{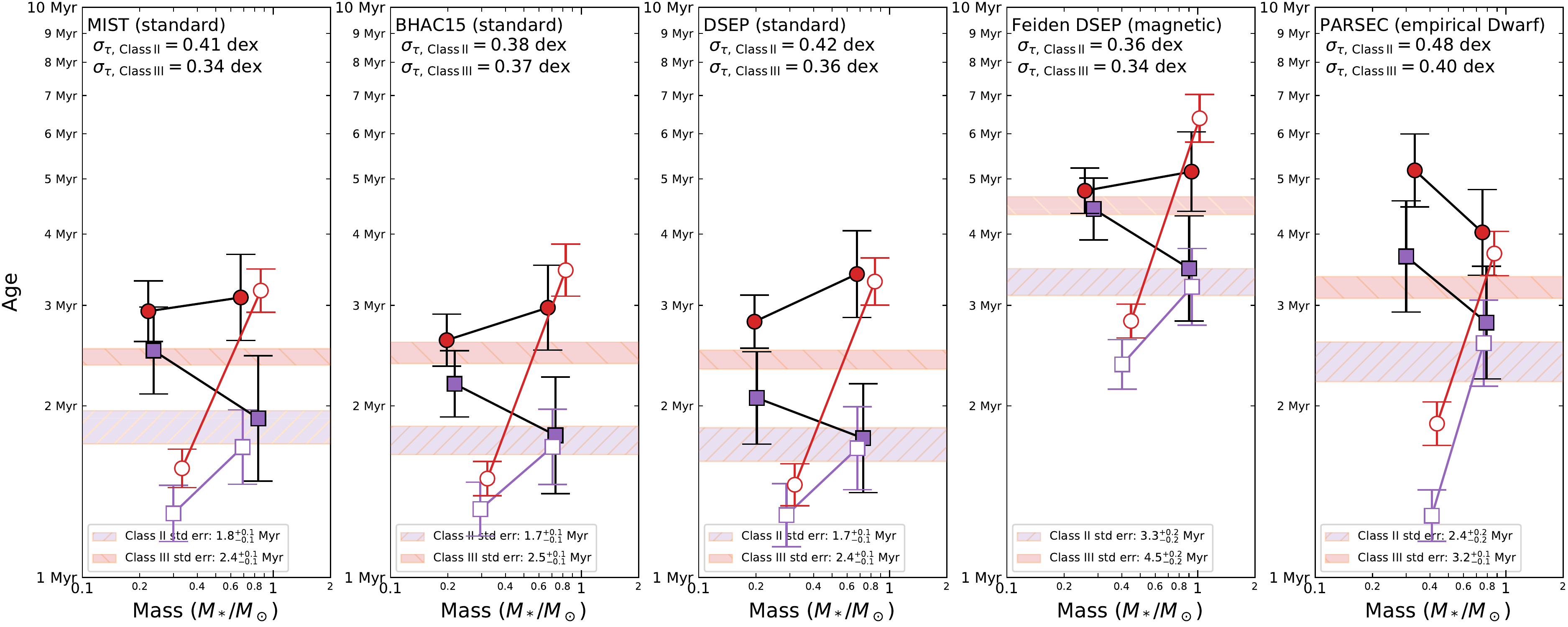}
\caption{Top: A comparison between inferred cluster ages with two mass bins between the \citet{2020ApJ...891...29S} models, using SPOTS $f=0$, $f=0.34$, and $f=0.5$---varying the starspot filling fraction. Bottom: A similar comparison for a variety of other evolutionary tracks, grouping by standard and non-standard tracks. The solid symbols are color-coded by class: Class II (purple) and Class III (red). The filled symbols correspond to the literature sample, and the unfilled symbols are the APOGEE Net SpT's. The symbols and error bars are the mean binned age and mass values and the dispersion of each bin, respectively. Bins are chosen for each plot by splitting the population into two equally populated groups of young stars across mass. Hatched bands correspond to the standard error of the population, and are represented in Table \ref{tab:ages}.} \label{fig:Figure_14}
\end{figure*}

There is an apparent difference in the derived ages for the Class II versus Class III populations, with Class II ages appearing systematically younger (see Table \ref{tab:ages}). Class II luminosities may be overestimated due to systematics in how we infer $A_V$ in Class II and III stars (see Sec. \ref{sec:ysoc_lit}).

Another interpretation for the difference in derived ages between our Class II \& III sources is that there is a significant age spread among members of {\lamori},
and the stars still retaining their disks as Class II objects are about 30\% younger than those which have already evolved to the Class III stage. However, the error bars on individual stellar ages are large enough that tighter constraints may be necessary to distinguish these populations in age.

\subsection{Age Spread} \label{sec:agespread}

\begin{figure}
\centering
\includegraphics[trim={4cm 0cm 4.5cm 0cm},clip,width=\columnwidth]{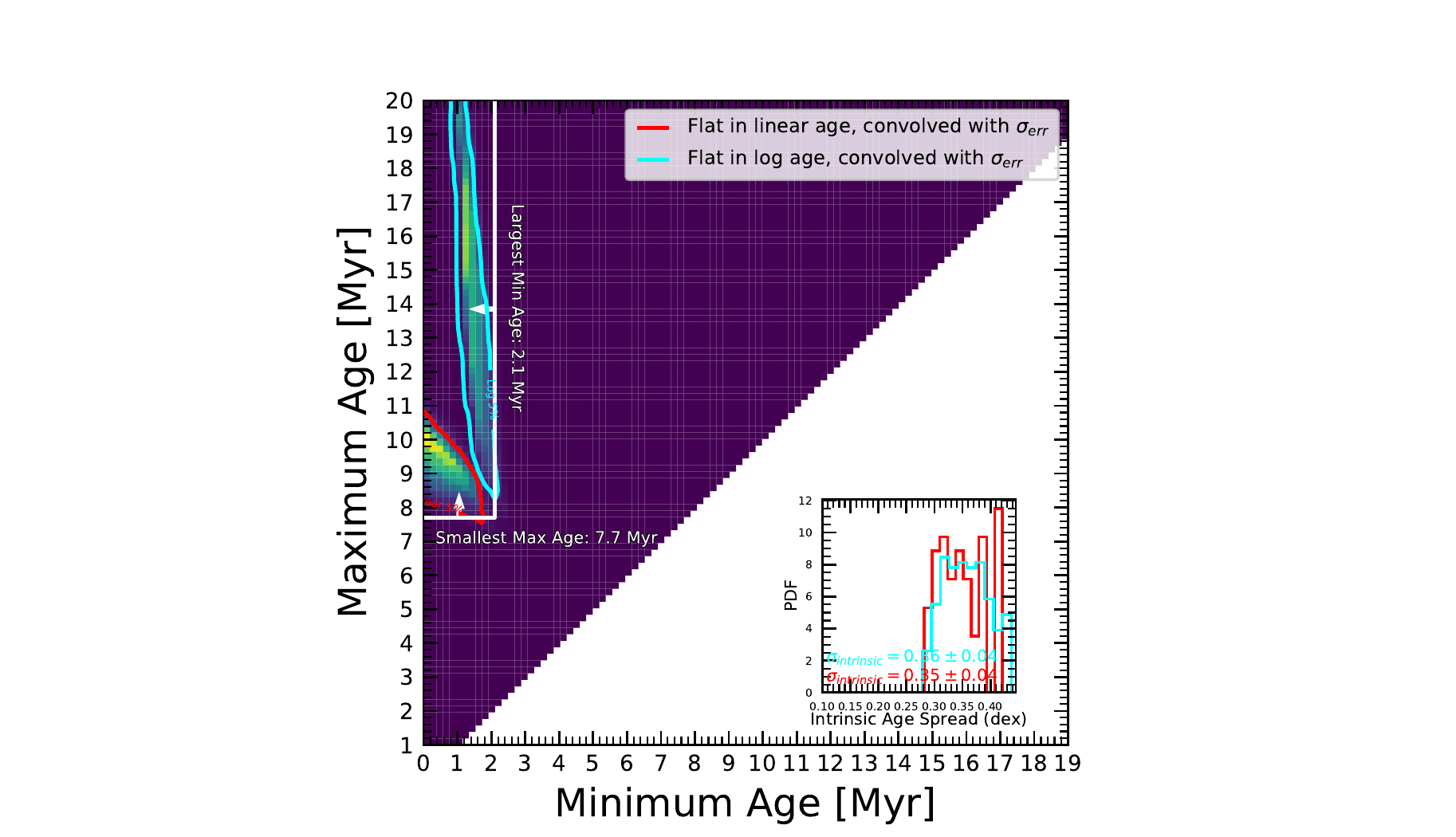}
\caption{Grid of maximum Kolmogorov-Schmidt p-values on distributions with flat ages in log and linear spaces ranging from the minimum age on the x-axis to the maximum age on the y-axis. Colors indicate an average of the Kolmogorov p-value with the flat log age distribution and the linear age distribution. Contours indicate the boundary for a KS p-value of $5\%$. Any distribution below the marked horizontal line, or to the right of the vertical line, is discrepant compared to the KS test at a power of $5\%$. These lines set the smallest choice of maximum age, and of the largest choice of minimum age which is indistinguishable from the observed age distribution. The inset plot is the calculated intrinsic age spread prior to a convolution with individual propagated observational age errors of the overlap region from the linear and logarithmic contours.} \label{fig:Figure_15}
\end{figure}

Even among kinematic cluster members, stellar ages in {\lamori} scatter depending on the class of star and the adopted evolutionary models.
Fig. \ref{fig:Figure_13} demonstrates that an assumption of a single extinction produces a significantly larger spread for the Class II sources than SED-fitting---some of the cluster members which are found to be heavily extincted young stars appear unphysically old (40---100 Myr). These erroneously old ages can largely be accounted for by considering the extinction around these Class II stars (right panels).
Binning these estimates of age produces a series of plots across spot filling fraction (top panels) and across different models (Fig. \ref{fig:Figure_14}). None of the panels using literature spectral types (filled markers) show a significant trend between low and high mass bins, which suggests that the population of available members in {\lamori} are too restricted in mass, or the temperature information is too uncertain, to analyze the slope of the HR diagram. It appears that the APOGEE Net stars (open markers) show a strong age trend with mass in all stars across all evolutionary models, but we do not attempt to interpret this behavior.

The SPOTS $f=0$ isochrones produce a similar result ($2.4 ^{+0.1}_{-0.1}$ Myr) to the other standard isochrones (MIST, BHAC15, DSEP). A comparison between the standard pre-MS DSEP isochrones and the Feiden magnetic DSEP isochrones also shows a similar systematic age offset from $2.4 ^{+0.1} _{-0.1}$ up to $4.5 ^{+0.2} _{-0.2}$ Myr. Combined with the analysis of the unperturbed stellar models from DSEP and SPOTS, this points to an answer of $2.5$ Myr for standard pre-MS isochrones, and a systematically older age ($4$--$5$ Myr) for the magnetic and spotted isochrones.

Through our analysis in {\lamori}, we find a relatively wide band of stellar sources spanning about $0.4$ dex in $\sigma _{\mathrm{log} L}$ and in the neighborhood of $0.4$ dex in $\sigma _{\mathrm{log} \tau}$ for both Class II and Class III sources. Despite the significant improvements in membership made by {\it Gaia} DR2, a robust observational spread in derived ages remains and does not appear to be easily mitigated by considering kinematics or regions (as in Fig. \ref{fig:Figure_12}).

In Fig. \ref{fig:Figure_15}, we investigate the age spread necessary to reproduce the scatter in age which we observe with our analysis. To do so, we simulate the underlying distribution of stellar ages using a grid of Kolmogorov-Schmidt tests. We use a Monte-Carlo method to perturb distributions drawn from a linearly and a logarithmically spaced age distribution, whose limits are given by its values along the x and y axis. Red and cyan contours correspond to trial age distributions that are spaced linearly or logarithmically, respectively. We convolve these distributions of stellar ages with propagated age errors from our prior work, and compare them to the derived ages from Table \ref{tab:stellarparams}. Any point within each of the red and cyan contours cannot be distinguished from the observed distribution at the 5\% level.

We find that any intrinsic age distribution that is consistent with our observed age distribution and inferred age errors \lamOri must include stars between the ages of $2.1$ and $7.7$ Myr. This means that there is a minimum age spread in the {\lamori} cluster which is consistent with our analysis, defined by a cluster age distribution ranging from $2.1$---$7.7$ Myr. This suggests that, in the absence of factors such as rotation or binarity, and within the propagated age uncertainties, the minimum intrinsic age spread that is consistent with our analysis is $\sigma _{\mathrm{log} \tau} \gtrsim 0.16$ dex, corresponding to a uniform age distribution spanning $2.1$---$7.7$ Myr. Including an assumption of $\sim$0.1 dex from simulations involving binaries \citep{2019A&A...623A.159P}, the minimum intrinsic age spread is $\gtrsim$0.19 dex.

The inset plot of Fig. \ref{fig:Figure_15} is the ``intrinsic age spread'', produced by calculating dispersions in the real distributions used in the KS test. As this is prior to the convolution of each distribution with the propagated age errors from our analysis, this represents the average true underlying distribution of ages that is consistent with data. The mean of the age spreads lies between $0.31$---$0.4$ dex, indistinguishable from the observed $0.4$ dex dispersion in ages.

\subsection{Robustness to systematics} \label{sec:systematics}

\begin{figure}
\centering
\includegraphics[trim={0cm 0cm 0cm 0cm},clip,width=\columnwidth]{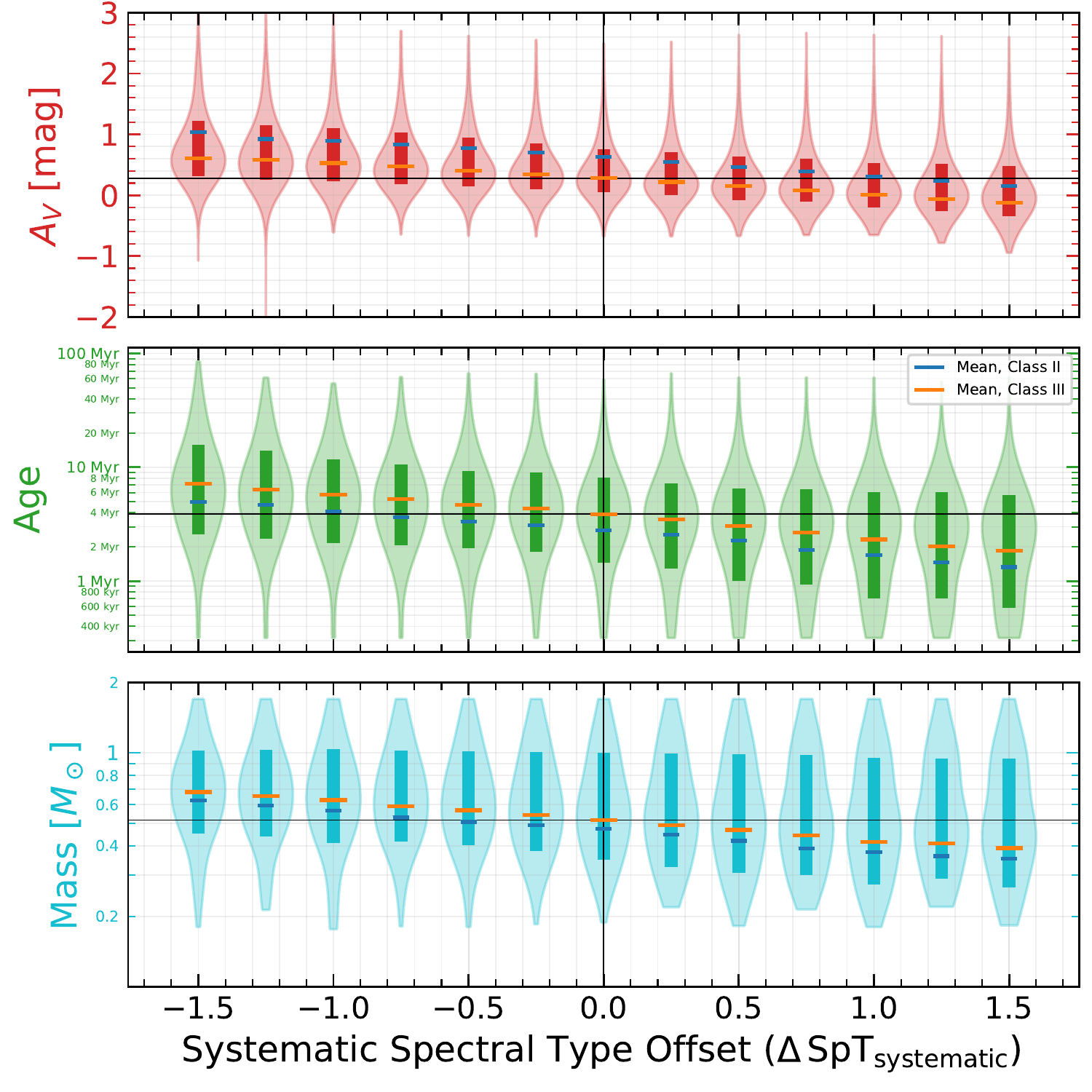}
\caption{Dependence of $A_V$, age, and mass distributions on systematic errors in spectral type in {\lamori}. Middle bars are means, and the shaded rectangular region is the calculated $1 \, \sigma$ dispersion in the entire population. Horizontal and vertical lines are drawn to guide the eye towards the mean parameters at 0 spectral type shift. Orange bars represent the Class III sources and blue bars represent the Class II sources.} \label{fig:Figure_16}
\end{figure}

It is possible that any global systematics in our extinction estimate, caused by a systematic in our input spectral types, might cause a systematic in the subsequent inferred ages and masses.
To probe the effects of systematics on possible inferred parameters for our cluster members, we run a grid of models in {\lamOri} and offset each spectral type by steps of $0.25$ subtypes in Fig. \ref{fig:Figure_16}.

Systematics in reported spectral type are not likely to be significant because the line of sight extinction to {\lamOri} is small, and because we have a physical constraint on our system---real $A_V$'s cannot be negative.
As a result, any systematic shift in temperature which results in negative extinctions for a significant number of the stars can be excluded. Only a systematic of $\pm 0.75$ subclasses may be consistent with a non-zero extinction. Prior studies that estimate the extinction in the neighborhood of {\lamori} have suggested cluster $A_V \sim 0.37$ mag \citep{1994ApJS...93..211D,2019A&A...628A.123Z}, which is consistent with a smaller systematic of $\pm 0.25$ subtypes. Because the variation in age and mass is relatively flat in this parameter space, this is an indication that our age determinations are relatively robust to reported temperature systematics.
The variation in SpT across $\pm 0.5$ subclasses corresponds to nearly $0.2$ mag, $1$ Myr, or $< 0.1 M_\odot$. This is likely representative of the order of any systematic error in our method.

\section{Discussion \& Conclusion} \label{sec:disc}
Motivated by the physical environment and youth of the {\lamori} association with improved {\it Gaia} membership information, we developed an open-cluster calibrated SED-fitting method applicable to young stars. We used external estimates of effective temperature in this cluster in the form of spectral types to constrain our extinction errors and derive a unique extinction for each star. The dereddened intrinsic colors from each star were uniquely determined by the tabulated colors from the \citet{2013ApJS..208....9P} young stellar table. This approach was appropriate for Class III (W1$-$W2$< 0.3$) stars but not for Class II (W1$-$W2$> 0.3$) stars, for which we needed to derive a single color that was less sensitive to the IR or UV excess which may not reflect photospheric colors in Classical T-Tauri stars.
We did not attempt to do full dust modelling of Class II sources. There is also no attempt to account for binarity, which may have biases on the age estimates of our population \citep{2006MNRAS.373.1251N} but may not significantly influence the width of the age distribution \citep{dolan2001}.

We demonstrate that the absolute age of the cluster varies from 2-3 Myr from theoretical non-spotted models to 4-5 Myr for magnetic and spotted models. This represents a factor of two difference in age determination for these young clusters that arises from non-traditional stellar physics. For stellar interiors models, we investigated modern calculations that include the effects of magnetic fields and starspots. We did not, however, include a range of spot filling factors or magnetic field strengths in our models, as a predictive framework for doing so is currently lacking.

Perhaps the best point of comparison for this work is in \citet{2018AJ....156...84K}, which used PHOENIX $T_{\mathrm{eff}}$'s from APOGEE spectra, bolometric corrections and intrinsic colors from \citet{1995ApJS..101..117K}, a dereddening process with {\it Gaia} BP-RP, and age estimation from their color-magnitude diagram or PARSEC-COLIBRI isochrones to estimate the ages of subgroups across the Orion star-forming complex. We also use $T_{\mathrm{eff}}$, color excess, and bolometric corrections as part of our dereddening process. In our treatment, we use empirical young colors, which appear to have a rather large offset compared to traditional dwarf colors (Fig. \ref{fig:Figure_9}), and we incorporate data from our Class II targets. Additionally, we use an SED method with a large set of open cluster calibrated colors on our weak-line T Tauri stars to infer $A_V$, and a set of similarly open cluster calibrated {\it Gaia} G-RP colors to infer $A_V$ for the photosphere of our Class II stars, reporting stellar parameters for all stars. Furthermore, we test multiple families of classical and magnetic/spotted isochrones in our Sec. \ref{sec:clusterage}.

The range of answers for \citet{2018AJ....156...84K}'s PARSEC-COLIBRI isochrones with {\lamOri} range from their systematically youngest subgroups ``near B30'' at $2.4 \pm 1.3$ Myr and ``near B35'' at $2.6 \pm 1.3$ Myr to their two core subgroups, $3.7 \pm 1.0$ Myr for the ``main'' and $5.1 \pm 1.1$ Myr for the nearby older subgroup. We find similar answers of $3.7 ^{+0.4}_{-0.2}$ Myr near B30, $2.2^{+0.6}_{-0.1}$ Myr near B35, and $4.5 ^{+0.6} _{-0.2}$ Myr for our population. Our kinematics also provide similar results; we clearly distinguish B30 and B35 from the central region, but we do not distinguish between LamOri-1 and LamOri-2, perhaps due to different membership critera. With systematics between the PHOENIX temperatures and our $T_{\mathrm{eff}}$'s, different intrinsic colors, and different dereddening procedures, these ages are roughly concordant. Where we find an extinction offset in common stars, this is due almost entirely due to the different temperature scales that we use; runs using their published temperatures yield extinctions with a small bias. Additionally, the age distributions, particularly the finding that B30 and B35 are not distinguishable from each other but are mutually distinguishable from the center, agrees with the result from \citet{dolan2001}.

Since starspot models do not affect age estimates in stars with radiative envelopes, spotted models infer older ages in low mass stars and agree with nonspotted models from higher mass stars. Nonmagnetic methods weighted towards higher mass stars give old ages, but largely provide ages that are too young in the low mass domain. The concordance between literature ages for {\lamOri} of $\sim$4-7 Myr and the magnetic method by our fiducial SPOTS model is encouraging; these previous ages \citep[e.g. by][]{2019A&A...628A.123Z} apply strong weights to high mass stars in the cluster, whereas our method weights all stars equally. Since there are significantly more lower mass members, this ends up being much more sensitive to the low-mass stars. Spots and magnetic effects applied to these low-mass stars appear to provide ages which agree well with the literature age of the cluster.
For future work, it could be useful to extend our starspot methodology across a much wider mass range, which would allow us to quantify the mass systematics and introduce a more robust age determination for the entire cluster.

We detected an observed spread in luminosities of $0.4$ dex, and observed spread in ages of $0.4$ dex. By doing an analysis on the likely underlying distribution of ages prior to convolution with our propagated age uncertainties, we suggest that the underlying intrinsic age spread of cluster members in {\lamori} is $\sim$0.35 dex in log age. This is consistent with the observational spread of $0.4$ dex throughout the Orion Complex. There are indications in the literature that the ONC may have a much tighter bound of $0.2$ dex in log age in works like \citet{2011MNRAS.418.1948J}. In our analysis, we find a minimum intrinsic age spread of $\sigma _{\mathrm{log} \tau} \gtrsim 0.16$ dex without considering binarity (and $0.19$ dex with a binary contribution of $0.1$ dex from \citet{2019A&A...623A.159P}) to be consistent with our propagated errors. We note that we appear to identify systematically larger extinctions (Fig. \ref{fig:Figure_7}) and therefore younger ages for our Class II population (Fig. \ref{fig:Figure_10} \& \ref{fig:Figure_13}), which is possible in a star-by-star extinction analysis, which may help explain the discrepancy between our findings. Also, in doing the analysis for the Class II members, we found that a single extinction value produces unphysically old ages of 40---100 Myr for this cluster where SED fitting techniques infer much more reasonable ages ($\lesssim$10 Myr) for most of these stars---this suggests that the Class II sources have a significant star-by-star variation in extinction (Fig. \ref{fig:Figure_13}).

However, there are a number of effects that could induce a range of $L$ even without an age spread: a distribution of spots/faculae or significant temperature uncertainties. We therefore caution that our (significant) minimum age range may be impacted by physical effects that are outside the scope of the current work. We do not attempt to do a spot inference model in {\lamori}, which precludes us from determining the true effect of the first possibility. Because spots and reddening can produce similar effects in colors but spots also affect evolution on the HRD, this is not a trivial effect. More precise temperatures are needed as an outsized fraction of the luminosity error is the large ($\sim$90$\%$) contribution from the uncertainty in spectral type. Disk effects may also affect the observational spread, but likely do not dominate, because of the significant luminosity spread of Class III stars. Other contributing factors of an age spread may include the bolometric correction varying with spot filling factor, systematics in the adopted empirical photospheric colors, and the effects of binaries. More work is needed to investigate the colors of pre-MS stars, including the effects of spots; such an effort would help further demonstrate whether the observed spread in luminosity corresponds to a large physical spread in ages.

Any systematics in spectral types are likely to be small because large systematics in spectral type would produce nonphysical mean extinctions. Consensus that this cluster has a mean extinction that is in the neighborhood of $\sim$0.4 mag is consistent with a strong constraint on the possible magnitude of any systematic to within $0.25$ subclasses. A weaker non-zero constraint on the extinctions provides a bound of $0.75$ subclasses. Throughout this range of possible systematics, the absolute age and inferred age spread using the SPOTS $f_{\mathrm{spot}}=0.34$ isochrones do not vary significantly, though the systematic age uncertainty may be on the order of $\pm 1$ Myr if typical temperature errors are representative of the size of global systematics.

While this work applied SED-fitting techniques to broadband photometry and analyzed the resultant age distributions with a wide number of different stellar evolution models, it is clear that still more robust kinematic analyses and spectro-photometry are necessary to reveal the properties of pre-MS stellar associations. Even in these young clusters, non-classical effects, such as magnetism and spots, greatly affect the absolute ages inferred in these clusters. These effects also influence trends in age derived by low and high mass members of the same cluster. Independent temperature measurements for a larger mass range of members will expand this analysis and allow more sensitive statistical tests to be performed to infer the extent of the age spread. From the modelling perspective, it continues to be a mystery whether a descriptive spot modelling method will help account for some of the star-by-star luminosity variations, particularly if a spread in spots or some other magnetic effect with rotation can reduce the required intrinsic age spread for this distribution.

This framework is presented as a package for the Young Stellar Object Corral (YSOC), where it is used to compute the extinction values and intrinsic stellar properties on all young stars where sufficient data is available.

\section*{Acknowledgements}
We thank Diego Godoy-Rivera for sharing open cluster membership data from {\it Gaia} in advance of publication. We thank Deokkeun An for sharing open cluster data products, as well as Marina Kounkel and Richard Olney for sharing relevant data products from APOGEE Net. We wish to thank Diego Godoy-Rivera and Tharindu Jayasinghe for useful discussions. We thank the anonymous referee for their extensive suggestions, which greatly improved this work. MHP and LC acknowledge support from NASA grant 80NSSC19K0597.

This work has made use of data from the European Space Agency (ESA) mission
{\it Gaia} (\url{https://www.cosmos.esa.int/gaia}), processed by the {\it Gaia}
Data Processing and Analysis Consortium (DPAC,
\url{https://www.cosmos.esa.int/web/gaia/dpac/consortium}). Funding for the DPAC
has been provided by national institutions, in particular the institutions
participating in the {\it Gaia} Multilateral Agreement.

\bibliography{cite}{}

\begin{thebibliography}{}
\expandafter\ifx\csname natexlab\endcsname\relax\def\natexlab#1{#1}\fi
\providecommand{\url}[1]{\href{#1}{#1}}
\providecommand{\dodoi}[1]{doi:~\href{http://doi.org/#1}{\nolinkurl{#1}}}
\providecommand{\doeprint}[1]{\href{http://ascl.net/#1}{\nolinkurl{http://ascl.net/#1}}}
\providecommand{\doarXiv}[1]{\href{https://arxiv.org/abs/#1}{\nolinkurl{https://arxiv.org/abs/#1}}}

\bibitem[{{Abolfathi} {et~al.}(2018){Abolfathi}, {Aguado}, {Aguilar}, {Allende
  Prieto}, {Almeida}, {Ananna}, {Anders}, {Anderson}, {Andrews}, {Anguiano},
  {Arag{\'o}n-Salamanca}, {Argudo-Fern{\'a}ndez}, {Armengaud}, {Ata},
  {Aubourg}, {Avila-Reese}, {Badenes}, {Bailey}, {Balland}, {Barger},
  {Barrera-Ballesteros}, {Bartosz}, {Bastien}, {Bates}, {Baumgarten},
  {Bautista}, {Beaton}, {Beers}, {Belfiore}, {Bender}, {Bernardi}, {Bershady},
  {Beutler}, {Bird}, {Bizyaev}, {Blanc}, {Blanton}, {Blomqvist}, {Bolton},
  {Boquien}, {Borissova}, {Bovy}, {Bradna Diaz}, {Brandt}, {Brinkmann},
  {Brownstein}, {Bundy}, {Burgasser}, {Burtin}, {Busca}, {Ca{\~n}as},
  {Cano-D{\'\i}az}, {Cappellari}, {Carrera}, {Casey}, {Cervantes Sodi}, {Chen},
  {Cherinka}, {Chiappini}, {Choi}, {Chojnowski}, {Chuang}, {Chung}, {Clerc},
  {Cohen}, {Comerford}, {Comparat}, {Correa do Nascimento}, {da Costa},
  {Cousinou}, {Covey}, {Crane}, {Cruz-Gonzalez}, {Cunha}, {da Silva Ilha},
  {Damke}, {Darling}, {Davidson}, {Dawson}, {de Icaza Lizaola}, {de la
  Macorra}, {de la Torre}, {De Lee}, {de Sainte Agathe}, {Deconto Machado},
  {Dell'Agli}, {Delubac}, {Diamond-Stanic}, {Donor}, {Downes}, {Drory}, {du Mas
  des Bourboux}, {Duckworth}, {Dwelly}, {Dyer}, {Ebelke}, {Davis Eigenbrot},
  {Eisenstein}, {Elsworth}, {Emsellem}, {Eracleous}, {Erfanianfar},
  {Escoffier}, {Fan}, {Fern{\'a}ndez Alvar}, {Fernandez-Trincado}, {Fernand o
  Cirolini}, {Feuillet}, {Finoguenov}, {Fleming}, {Font-Ribera}, {Freischlad},
  {Frinchaboy}, {Fu}, {G{\'o}mez Maqueo Chew}, {Galbany}, {Garc{\'\i}a
  P{\'e}rez}, {Garcia-Dias}, {Garc{\'\i}a-Hern{\'a}ndez}, {Garma Oehmichen},
  {Gaulme}, {Gelfand }, {Gil-Mar{\'\i}n}, {Gillespie}, {Goddard}, {Gonz{\'a}lez
  Hern{\'a}ndez}, {Gonzalez-Perez}, {Grabowski}, {Green}, {Grier}, {Gueguen},
  {Guo}, {Guy}, {Hagen}, {Hall}, {Harding}, {Hasselquist}, {Hawley}, {Hayes},
  {Hearty}, {Hekker}, {Hernand ez}, {Hernandez Toledo}, {Hogg},
  {Holley-Bockelmann}, {Holtzman}, {Hou}, {Hsieh}, {Hunt}, {Hutchinson},
  {Hwang}, {Jimenez Angel}, {Johnson}, {Jones}, {J{\"o}nsson}, {Jullo}, {Khan},
  {Kinemuchi}, {Kirkby}, {Kirkpatrick}, {Kitaura}, {Knapp}, {Kneib},
  {Kollmeier}, {Lacerna}, {Lane}, {Lang}, {Law}, {Le Goff}, {Lee}, {Li}, {Li},
  {Lian}, {Liang}, {Lima}, {Lin}, {Long}, {Lucatello}, {Lundgren}, {Mackereth},
  {MacLeod}, {Mahadevan}, {Maia}, {Majewski}, {Manchado}, {Maraston},
  {Mariappan}, {Marques-Chaves}, {Masseron}, {Masters}, {McDermid}, {McGreer},
  {Melendez}, {Meneses-Goytia}, {Merloni}, {Merrifield}, {Meszaros}, {Meza},
  {Minchev}, {Minniti}, {Mueller}, {Muller-Sanchez}, {Muna}, {Mu{\~n}oz},
  {Myers}, {Nair}, {Nand ra}, {Ness}, {Newman}, {Nichol}, {Nidever},
  {Nitschelm}, {Noterdaeme}, {O'Connell}, {Oelkers}, {Oravetz}, {Oravetz},
  {Ort{\'\i}z}, {Osorio}, {Pace}, {Padilla}, {Palanque-Delabrouille},
  {Palicio}, {Pan}, {Pan}, {Parikh}, {P{\^a}ris}, {Park}, {Peirani},
  {Pellejero-Ibanez}, {Penny}, {Percival}, {Perez-Fournon}, {Petitjean},
  {Pieri}, {Pinsonneault}, {Pisani}, {Prada}, {Prakash}, {Queiroz}, {Raddick},
  {Raichoor}, {Barboza Rembold}, {Richstein}, {Riffel}, {Riffel}, {Rix},
  {Robin}, {Rodr{\'\i}guez Torres}, {Rom{\'a}n-Z{\'u}{\~n}iga}, {Ross},
  {Rossi}, {Ruan}, {Ruggeri}, {Ruiz}, {Salvato}, {S{\'a}nchez}, {S{\'a}nchez},
  {Sanchez Almeida}, {S{\'a}nchez-Gallego}, {Santana Rojas}, {Santiago},
  {Schiavon}, {Schimoia}, {Schlafly}, {Schlegel}, {Schneider}, {Schuster},
  {Schwope}, {Seo}, {Serenelli}, {Shen}, {Shen}, {Shetrone}, {Shull}, {Silva
  Aguirre}, {Simon}, {Skrutskie}, {Slosar}, {Smethurst}, {Smith}, {Sobeck},
  {Somers}, {Souter}, {Souto}, {Spindler}, {Stark}, {Stassun}, {Steinmetz},
  {Stello}, {Storchi-Bergmann}, {Streblyanska}, {Stringfellow}, {Su{\'a}rez},
  {Sun}, {Szigeti}, {Taghizadeh-Popp}, {Talbot}, {Tang}, {Tao}, {Tayar},
  {Tembe}, {Teske}, {Thakar}, {Thomas}, {Tissera}, {Tojeiro}, {Tremonti},
  {Troup}, {Urry}, {Valenzuela}, {van den Bosch}, {Vargas-Gonz{\'a}lez},
  {Vargas-Maga{\~n}a}, {Vazquez}, {Villanova}, {Vogt}, {Wake}, {Wang},
  {Weaver}, {Weijmans}, {Weinberg}, {Westfall}, {Whelan}, {Wilcots}, {Wild},
  {Williams}, {Wilson}, {Wood-Vasey}, {Wylezalek}, {Xiao}, {Yan}, {Yang},
  {Ybarra}, {Y{\`e}che}, {Zakamska}, {Zamora}, {Zarrouk}, {Zasowski}, {Zhang},
  {Zhao}, {Zhao}, {Zheng}, {Zheng}, {Zhou}, {Zhu}, {Zinn}, \&
  {Zou}}]{2018ApJS..235...42A}
{Abolfathi}, B., {Aguado}, D.~S., {Aguilar}, G., {et~al.} 2018, \apjs, 235, 42,
  \dodoi{10.3847/1538-4365/aa9e8a}

\bibitem[{{An} {et~al.}(2007){An}, {Terndrup}, {Pinsonneault}, {Paulson},
  {Hanson}, \& {Stauffer}}]{2007ApJ...655..233A}
{An}, D., {Terndrup}, D.~M., {Pinsonneault}, M.~H., {et~al.} 2007, \apj, 655,
  233, \dodoi{10.1086/509653}

\bibitem[{{Bailer-Jones}(2011)}]{2011MNRAS.411..435B}
{Bailer-Jones}, C.~A.~L. 2011, \mnras, 411, 435,
  \dodoi{10.1111/j.1365-2966.2010.17699.x}

\bibitem[{{Ballesteros-Paredes} \& {Hartmann}(2007)}]{2007RMxAA..43..123B}
{Ballesteros-Paredes}, J., \& {Hartmann}, L. 2007, \rmxaa, 43, 123

\bibitem[{{Baraffe} {et~al.}(2015){Baraffe}, {Homeier}, {Allard}, \&
  {Chabrier}}]{2015A&A...577A..42B}
{Baraffe}, I., {Homeier}, D., {Allard}, F., \& {Chabrier}, G. 2015, \aap, 577,
  A42, \dodoi{10.1051/0004-6361/201425481}

\bibitem[{{Barrado y Navascu{\'e}s} {et~al.}(2007){Barrado y Navascu{\'e}s},
  {Stauffer}, {Morales-Calder{\'o}n}, {Bayo}, {Fazzio}, {Megeath}, {Allen},
  {Hartmann}, \& {Calvet}}]{2007ApJ...664..481B}
{Barrado y Navascu{\'e}s}, D., {Stauffer}, J.~R., {Morales-Calder{\'o}n}, M.,
  {et~al.} 2007, \apj, 664, 481, \dodoi{10.1086/518816}

\bibitem[{{Bayo} {et~al.}(2012){Bayo}, {Barrado}, {Hu{\'e}lamo},
  {Morales-Calder{\'o}n}, {Melo}, {Stauffer}, \&
  {Stelzer}}]{2012A&A...547A..80B}
{Bayo}, A., {Barrado}, D., {Hu{\'e}lamo}, N., {et~al.} 2012, \aap, 547, A80,
  \dodoi{10.1051/0004-6361/201219374}

\bibitem[{{Bayo} {et~al.}(2008){Bayo}, {Rodrigo}, {Barrado Y Navascu{\'e}s},
  {Solano}, {Guti{\'e}rrez}, {Morales-Calder{\'o}n}, \& {Allard}}]{bayo2008}
{Bayo}, A., {Rodrigo}, C., {Barrado Y Navascu{\'e}s}, D., {et~al.} 2008, \aap,
  492, 277, \dodoi{10.1051/0004-6361:200810395}

\bibitem[{{Bayo} {et~al.}(2011){Bayo}, {Barrado}, {Stauffer},
  {Morales-Calder{\'o}n}, {Melo}, {Hu{\'e}lamo}, {Bouy}, {Stelzer}, {Tamura},
  \& {Jayawardhana}}]{2011A&A...536A..63B}
{Bayo}, A., {Barrado}, D., {Stauffer}, J., {et~al.} 2011, \aap, 536, A63,
  \dodoi{10.1051/0004-6361/201116617}

\bibitem[{{Bressan} {et~al.}(2012){Bressan}, {Marigo}, {Girardi}, {Salasnich},
  {Dal Cero}, {Rubele}, \& {Nanni}}]{2012MNRAS.427..127B}
{Bressan}, A., {Marigo}, P., {Girardi}, L., {et~al.} 2012, \mnras, 427, 127,
  \dodoi{10.1111/j.1365-2966.2012.21948.x}

\bibitem[{{Chabrier}(2003)}]{2003PASP..115..763C}
{Chabrier}, G. 2003, \pasp, 115, 763, \dodoi{10.1086/376392}

\bibitem[{{Chambers} {et~al.}(2016){Chambers}, {Magnier}, {Metcalfe},
  {Flewelling}, {Huber}, {Waters}, {Denneau}, {Draper}, {Farrow}, {Finkbeiner},
  {Holmberg}, {Koppenhoefer}, {Price}, {Rest}, {Saglia}, {Schlafly}, {Smartt},
  {Sweeney}, {Wainscoat}, {Burgett}, {Chastel}, {Grav}, {Heasley}, {Hodapp},
  {Jedicke}, {Kaiser}, {Kudritzki}, {Luppino}, {Lupton}, {Monet}, {Morgan},
  {Onaka}, {Shiao}, {Stubbs}, {Tonry}, {White}, {Ba{\~n}ados}, {Bell},
  {Bender}, {Bernard}, {Boegner}, {Boffi}, {Botticella}, {Calamida},
  {Casertano}, {Chen}, {Chen}, {Cole}, {Deacon}, {Frenk}, {Fitzsimmons},
  {Gezari}, {Gibbs}, {Goessl}, {Goggia}, {Gourgue}, {Goldman}, {Grant},
  {Grebel}, {Hambly}, {Hasinger}, {Heavens}, {Heckman}, {Henderson}, {Henning},
  {Holman}, {Hopp}, {Ip}, {Isani}, {Jackson}, {Keyes}, {Koekemoer}, {Kotak},
  {Le}, {Liska}, {Long}, {Lucey}, {Liu}, {Martin}, {Masci}, {McLean}, {Mindel},
  {Misra}, {Morganson}, {Murphy}, {Obaika}, {Narayan}, {Nieto-Santisteban},
  {Norberg}, {Peacock}, {Pier}, {Postman}, {Primak}, {Rae}, {Rai}, {Riess},
  {Riffeser}, {Rix}, {R{\"o}ser}, {Russel}, {Rutz}, {Schilbach}, {Schultz},
  {Scolnic}, {Strolger}, {Szalay}, {Seitz}, {Small}, {Smith}, {Soderblom},
  {Taylor}, {Thomson}, {Taylor}, {Thakar}, {Thiel}, {Thilker}, {Unger},
  {Urata}, {Valenti}, {Wagner}, {Walder}, {Walter}, {Watters}, {Werner},
  {Wood-Vasey}, \& {Wyse}}]{2016arXiv161205560C}
{Chambers}, K.~C., {Magnier}, E.~A., {Metcalfe}, N., {et~al.} 2016, arXiv
  e-prints, arXiv:1612.05560.
\newblock \doarXiv{1612.05560}

\bibitem[{{Chen} {et~al.}(2015){Chen}, {Bressan}, {Girardi}, {Marigo}, {Kong},
  \& {Lanza}}]{2015MNRAS.452.1068C}
{Chen}, Y., {Bressan}, A., {Girardi}, L., {et~al.} 2015, \mnras, 452, 1068,
  \dodoi{10.1093/mnras/stv1281}

\bibitem[{{Choi} {et~al.}(2016){Choi}, {Dotter}, {Conroy}, {Cantiello},
  {Paxton}, \& {Johnson}}]{2016ApJ...823..102C}
{Choi}, J., {Dotter}, A., {Conroy}, C., {et~al.} 2016, \apj, 823, 102,
  \dodoi{10.3847/0004-637X/823/2/102}

\bibitem[{{Cottle} {et~al.}(2018){Cottle}, {Covey}, {Suarez}, {Roman-Zuniga},
  {Schlafly}, {Downes}, {Ybarra}, {Hernand ez}, {Stassun}, {Stringfellow},
  {Getman}, {Feigelson}, {Borissova}, {Kim}, {Roman-Lopes}, {da Rio}, {de},
  {Frinchaboy}, {Kounkel}, {Majewski}, {Mennickent}, {Nidever}, {Nitschelm},
  {Pan}, {Shetrone}, {Zasowski}, {Chambers}, {Magnier}, \&
  {Valenti}}]{2018yCat..22360027C}
{Cottle}, J., {Covey}, K.~R., {Suarez}, G., {et~al.} 2018, VizieR Online Data
  Catalog, J/ApJS/236/27

\bibitem[{{Cutri} {et~al.}(2003){Cutri}, {Skrutskie}, {van Dyk}, {Beichman},
  {Carpenter}, {Chester}, {Cambresy}, {Evans}, {Fowler}, {Gizis}, {Howard},
  {Huchra}, {Jarrett}, {Kopan}, {Kirkpatrick}, {Light}, {Marsh}, {McCallon},
  {Schneider}, {Stiening}, {Sykes}, {Weinberg}, {Wheaton}, {Wheelock}, \&
  {Zacarias}}]{cutri2003}
{Cutri}, R.~M., {Skrutskie}, M.~F., {van Dyk}, S., {et~al.} 2003, {2MASS All
  Sky Catalog of point sources.}

\bibitem[{{Da Rio} {et~al.}(2010){Da Rio}, {Robberto}, {Soderblom}, {Panagia},
  {Hillenbrand}, {Palla}, \& {Stassun}}]{2010ApJ...722.1092D}
{Da Rio}, N., {Robberto}, M., {Soderblom}, D.~R., {et~al.} 2010, \apj, 722,
  1092, \dodoi{10.1088/0004-637X/722/2/1092}

\bibitem[{{David} {et~al.}(2019){David}, {Hillenbrand}, {Gillen}, {Cody},
  {Howell}, {Isaacson}, \& {Livingston}}]{2019ApJ...872..161D}
{David}, T.~J., {Hillenbrand}, L.~A., {Gillen}, E., {et~al.} 2019, \apj, 872,
  161, \dodoi{10.3847/1538-4357/aafe09}

\bibitem[{{Davies}(2015)}]{2015PhDT.......219D}
{Davies}, C.~L. 2015, PhD thesis, University of St.~Andrews

\bibitem[{{Diplas} \& {Savage}(1994)}]{1994ApJS...93..211D}
{Diplas}, A., \& {Savage}, B.~D. 1994, \apjs, 93, 211, \dodoi{10.1086/192052}

\bibitem[{{Dolan} \& {Mathieu}(2001)}]{dolan2001}
{Dolan}, C.~J., \& {Mathieu}, R.~D. 2001, \aj, 121, 2124,
  \dodoi{10.1086/319946}

\bibitem[{{Dolan} \& {Mathieu}(2002)}]{dolan2002}
---. 2002, \aj, 123, 387, \dodoi{10.1086/324631}

\bibitem[{{Dotter} {et~al.}(2008){Dotter}, {Chaboyer}, {Jevremovi{\'c}},
  {Kostov}, {Baron}, \& {Ferguson}}]{2008ApJS..178...89D}
{Dotter}, A., {Chaboyer}, B., {Jevremovi{\'c}}, D., {et~al.} 2008, \apjs, 178,
  89, \dodoi{10.1086/589654}

\bibitem[{{Elmegreen}(2007)}]{2007ApJ...668.1064E}
{Elmegreen}, B.~G. 2007, \apj, 668, 1064, \dodoi{10.1086/521327}

\bibitem[{{Fang} {et~al.}(2017){Fang}, {Kim}, {Pascucci}, {Apai}, {Zhang},
  {Sicilia-Aguilar}, {Alonso-Mart{\'{\i}}nez}, {Eiroa}, \&
  {Wang}}]{2017AJ....153..188F}
{Fang}, M., {Kim}, J.~S., {Pascucci}, I., {et~al.} 2017, \aj, 153, 188,
  \dodoi{10.3847/1538-3881/aa647b}

\bibitem[{{Feiden}(2016)}]{2016A&A...593A..99F}
{Feiden}, G.~A. 2016, \aap, 593, A99, \dodoi{10.1051/0004-6361/201527613}

\bibitem[{{Feiden} \& {Chaboyer}(2012)}]{2012ApJ...761...30F}
{Feiden}, G.~A., \& {Chaboyer}, B. 2012, \apj, 761, 30,
  \dodoi{10.1088/0004-637X/761/1/30}

\bibitem[{{Feiden} \& {Chaboyer}(2014)}]{2014ApJ...789...53F}
---. 2014, \apj, 789, 53, \dodoi{10.1088/0004-637X/789/1/53}

\bibitem[{{Fiorucci} \& {Munari}(2003)}]{2003A&A...401..781F}
{Fiorucci}, M., \& {Munari}, U. 2003, \aap, 401, 781,
  \dodoi{10.1051/0004-6361:20030075}

\bibitem[{{Flewelling} {et~al.}(2016){Flewelling}, {Magnier}, {Chambers},
  {Heasley}, {Holmberg}, {Huber}, {Sweeney}, {Waters}, {Chen}, {Farrow},
  {Hasinger}, {Henderson}, {Long}, {Metcalfe}, {Nieto-Santisteban}, {Norberg},
  {Saglia}, {Szalay}, {Rest}, {Thakar}, {Tonry}, {Valenti}, {Werner}, {White},
  {Denneau}, {Draper}, {Hodapp}, {Jedicke}, {Kaiser}, {Kudritzki}, {Price},
  {Wainscoat}, {Chastel}, {McClean}, {Postman}, \& {Shiao}}]{flewelling2016}
{Flewelling}, H.~A., {Magnier}, E.~A., {Chambers}, K.~C., {et~al.} 2016, arXiv
  e-prints, arXiv:1612.05243.
\newblock \doarXiv{1612.05243}

\bibitem[{{Gaia Collaboration} {et~al.}(2018){Gaia Collaboration}, {Brown},
  {Vallenari}, {Prusti}, {de Bruijne}, {Babusiaux}, {Bailer-Jones}, {Biermann},
  {Evans}, {Eyer}, {Jansen}, {Jordi}, {Klioner}, {Lammers}, {Lindegren},
  {Luri}, {Mignard}, {Panem}, {Pourbaix}, {Randich}, {Sartoretti}, {Siddiqui},
  {Soubiran}, {van Leeuwen}, {Walton}, {Arenou}, {Bastian}, {Cropper},
  {Drimmel}, {Katz}, {Lattanzi}, {Bakker}, {Cacciari}, {Casta{\~n}eda},
  {Chaoul}, {Cheek}, {De Angeli}, {Fabricius}, {Guerra}, {Holl}, {Masana},
  {Messineo}, {Mowlavi}, {Nienartowicz}, {Panuzzo}, {Portell}, {Riello},
  {Seabroke}, {Tanga}, {Th{\'e}venin}, {Gracia-Abril}, {Comoretto},
  {Garcia-Reinaldos}, {Teyssier}, {Altmann}, {Andrae}, {Audard},
  {Bellas-Velidis}, {Benson}, {Berthier}, {Blomme}, {Burgess}, {Busso},
  {Carry}, {Cellino}, {Clementini}, {Clotet}, {Creevey}, {Davidson}, {De
  Ridder}, {Delchambre}, {Dell'Oro}, {Ducourant},
  {Fern{\'a}ndez-Hern{\'a}ndez}, {Fouesneau}, {Fr{\'e}mat}, {Galluccio},
  {Garc{\'\i}a-Torres}, {Gonz{\'a}lez-N{\'u}{\~n}ez}, {Gonz{\'a}lez-Vidal},
  {Gosset}, {Guy}, {Halbwachs}, {Hambly}, {Harrison}, {Hern{\'a}ndez},
  {Hestroffer}, {Hodgkin}, {Hutton}, {Jasniewicz}, {Jean-Antoine-Piccolo},
  {Jordan}, {Korn}, {Krone-Martins}, {Lanzafame}, {Lebzelter}, {L{\"o}ffler},
  {Manteiga}, {Marrese}, {Mart{\'\i}n-Fleitas}, {Moitinho}, {Mora}, {Muinonen},
  {Osinde}, {Pancino}, {Pauwels}, {Petit}, {Recio-Blanco}, {Richards},
  {Rimoldini}, {Robin}, {Sarro}, {Siopis}, {Smith}, {Sozzetti}, {S{\"u}veges},
  {Torra}, {van Reeven}, {Abbas}, {Abreu Aramburu}, {Accart}, {Aerts},
  {Altavilla}, {{\'A}lvarez}, {Alvarez}, {Alves}, {Anderson}, {Andrei},
  {Anglada Varela}, {Antiche}, {Antoja}, {Arcay}, {Astraatmadja}, {Bach},
  {Baker}, {Balaguer-N{\'u}{\~n}ez}, {Balm}, {Barache}, {Barata}, {Barbato},
  {Barblan}, {Barklem}, {Barrado}, {Barros}, {Barstow}, {Bartholom{\'e}
  Mu{\~n}oz}, {Bassilana}, {Becciani}, {Bellazzini}, {Berihuete}, {Bertone},
  {Bianchi}, {Bienaym{\'e}}, {Blanco-Cuaresma}, {Boch}, {Boeche}, {Bombrun},
  {Borrachero}, {Bossini}, {Bouquillon}, {Bourda}, {Bragaglia}, {Bramante},
  {Breddels}, {Bressan}, {Brouillet}, {Br{\"u}semeister}, {Brugaletta},
  {Bucciarelli}, {Burlacu}, {Busonero}, {Butkevich}, {Buzzi}, {Caffau},
  {Cancelliere}, {Cannizzaro}, {Cantat-Gaudin}, {Carballo}, {Carlucci},
  {Carrasco}, {Casamiquela}, {Castellani}, {Castro-Ginard}, {Charlot},
  {Chemin}, {Chiavassa}, {Cocozza}, {Costigan}, {Cowell}, {Crifo}, {Crosta},
  {Crowley}, {Cuypers}, {Dafonte}, {Damerdji}, {Dapergolas}, {David}, {David},
  {de Laverny}, {De Luise}, {De March}, {de Martino}, {de Souza}, {de Torres},
  {Debosscher}, {del Pozo}, {Delbo}, {Delgado}, {Delgado}, {Di Matteo},
  {Diakite}, {Diener}, {Distefano}, {Dolding}, {Drazinos}, {Dur{\'a}n},
  {Edvardsson}, {Enke}, {Eriksson}, {Esquej}, {Eynard Bontemps}, {Fabre},
  {Fabrizio}, {Faigler}, {Falc{\~a}o}, {Farr{\`a}s Casas}, {Federici},
  {Fedorets}, {Fernique}, {Figueras}, {Filippi}, {Findeisen}, {Fonti},
  {Fraile}, {Fraser}, {Fr{\'e}zouls}, {Gai}, {Galleti}, {Garabato},
  {Garc{\'\i}a-Sedano}, {Garofalo}, {Garralda}, {Gavel}, {Gavras}, {Gerssen},
  {Geyer}, {Giacobbe}, {Gilmore}, {Girona}, {Giuffrida}, {Glass}, {Gomes},
  {Granvik}, {Gueguen}, {Guerrier}, {Guiraud}, {Guti{\'e}rrez-S{\'a}nchez},
  {Haigron}, {Hatzidimitriou}, {Hauser}, {Haywood}, {Heiter}, {Helmi}, {Heu},
  {Hilger}, {Hobbs}, {Hofmann}, {Holland}, {Huckle}, {Hypki}, {Icardi},
  {Jan{\ss}en}, {Jevardat de Fombelle}, {Jonker}, {Juh{\'a}sz}, {Julbe},
  {Karampelas}, {Kewley}, {Klar}, {Kochoska}, {Kohley}, {Kolenberg},
  {Kontizas}, {Kontizas}, {Koposov}, {Kordopatis}, {Kostrzewa-Rutkowska},
  {Koubsky}, {Lambert}, {Lanza}, {Lasne}, {Lavigne}, {Le Fustec}, {Le
  Poncin-Lafitte}, {Lebreton}, {Leccia}, {Leclerc}, {Lecoeur-Taibi},
  {Lenhardt}, {Leroux}, {Liao}, {Licata}, {Lindstr{\o}m}, {Lister}, {Livanou},
  {Lobel}, {L{\'o}pez}, {Managau}, {Mann}, {Mantelet}, {Marchal}, {Marchant},
  {Marconi}, {Marinoni}, {Marschalk{\'o}}, {Marshall}, {Martino}, {Marton},
  {Mary}, {Massari}, {Matijevi{\v{c}}}, {Mazeh}, {McMillan}, {Messina},
  {Michalik}, {Millar}, {Molina}, {Molinaro}, {Moln{\'a}r}, {Montegriffo},
  {Mor}, {Morbidelli}, {Morel}, {Morris}, {Mulone}, {Muraveva}, {Musella},
  {Nelemans}, {Nicastro}, {Noval}, {O'Mullane}, {Ord{\'e}novic},
  {Ord{\'o}{\~n}ez-Blanco}, {Osborne}, {Pagani}, {Pagano}, {Pailler},
  {Palacin}, {Palaversa}, {Panahi}, {Pawlak}, {Piersimoni}, {Pineau}, {Plachy},
  {Plum}, {Poggio}, {Poujoulet}, {Pr{\v{s}}a}, {Pulone}, {Racero}, {Ragaini},
  {Rambaux}, {Ramos-Lerate}, {Regibo}, {Reyl{\'e}}, {Riclet}, {Ripepi}, {Riva},
  {Rivard}, {Rixon}, {Roegiers}, {Roelens}, {Romero-G{\'o}mez}, {Rowell},
  {Royer}, {Ruiz-Dern}, {Sadowski}, {Sagrist{\`a} Sell{\'e}s}, {Sahlmann},
  {Salgado}, {Salguero}, {Sanna}, {Santana-Ros}, {Sarasso}, {Savietto},
  {Schultheis}, {Sciacca}, {Segol}, {Segovia}, {S{\'e}gransan}, {Shih},
  {Siltala}, {Silva}, {Smart}, {Smith}, {Solano}, {Solitro}, {Sordo}, {Soria
  Nieto}, {Souchay}, {Spagna}, {Spoto}, {Stampa}, {Steele},
  {Steidelm{\"u}ller}, {Stephenson}, {Stoev}, {Suess}, {Surdej}, {Szabados},
  {Szegedi-Elek}, {Tapiador}, {Taris}, {Tauran}, {Taylor}, {Teixeira},
  {Terrett}, {Teyssand ier}, {Thuillot}, {Titarenko}, {Torra Clotet}, {Turon},
  {Ulla}, {Utrilla}, {Uzzi}, {Vaillant}, {Valentini}, {Valette}, {van Elteren},
  {Van Hemelryck}, {van Leeuwen}, {Vaschetto}, {Vecchiato}, {Veljanoski},
  {Viala}, {Vicente}, {Vogt}, {von Essen}, {Voss}, {Votruba}, {Voutsinas},
  {Walmsley}, {Weiler}, {Wertz}, {Wevers}, {Wyrzykowski}, {Yoldas},
  {{\v{Z}}erjal}, {Ziaeepour}, {Zorec}, {Zschocke}, {Zucker}, {Zurbach}, \&
  {Zwitter}}]{2018A&A...616A...1G}
{Gaia Collaboration}, {Brown}, A.~G.~A., {Vallenari}, A., {et~al.} 2018, \aap,
  616, A1, \dodoi{10.1051/0004-6361/201833051}

\bibitem[{{Gao}(2018)}]{2018PASP..130l4101G}
{Gao}, X.-h. 2018, \pasp, 130, 124101, \dodoi{10.1088/1538-3873/aae0d2}

\bibitem[{{Godoy-Rivera} {et~al.}(2021){Godoy-Rivera}, {Pinsonneault}, \&
  {Rebull}}]{2021arXiv210101183G}
{Godoy-Rivera}, D., {Pinsonneault}, M.~H., \& {Rebull}, L.~M. 2021, arXiv
  e-prints, arXiv:2101.01183.
\newblock \doarXiv{2101.01183}

\bibitem[{{Gossage} {et~al.}(2018){Gossage}, {Conroy}, {Dotter}, {Choi},
  {Rosenfield}, {Cargile}, \& {Dolphin}}]{2018ApJ...863...67G}
{Gossage}, S., {Conroy}, C., {Dotter}, A., {et~al.} 2018, \apj, 863, 67,
  \dodoi{10.3847/1538-4357/aad0a0}

\bibitem[{{Gully-Santiago} {et~al.}(2017){Gully-Santiago}, {Herczeg},
  {Czekala}, {Somers}, {Grankin}, {Covey}, {Donati}, {Alencar}, {Hussain},
  {Shappee}, {Mace}, {Lee}, {Holoien}, {Jose}, \& {Liu}}]{2017ApJ...836..200G}
{Gully-Santiago}, M.~A., {Herczeg}, G.~J., {Czekala}, I., {et~al.} 2017, \apj,
  836, 200, \dodoi{10.3847/1538-4357/836/2/200}

\bibitem[{{Henden} {et~al.}(2016){Henden}, {Templeton}, {Terrell}, {Smith},
  {Levine}, \& {Welch}}]{2016yCat.2336....0H}
{Henden}, A.~A., {Templeton}, M., {Terrell}, D., {et~al.} 2016, VizieR Online
  Data Catalog, II/336

\bibitem[{{Hern{\'a}ndez} {et~al.}(2010){Hern{\'a}ndez}, {Morales-Calderon},
  {Calvet}, {Hartmann}, {Muzerolle}, {Gutermuth}, {Luhman}, \&
  {Stauffer}}]{2010ApJ...722.1226H}
{Hern{\'a}ndez}, J., {Morales-Calderon}, M., {Calvet}, N., {et~al.} 2010, \apj,
  722, 1226, \dodoi{10.1088/0004-637X/722/2/1226}

\bibitem[{{Hillenbrand}(1997)}]{1997AJ....113.1733H}
{Hillenbrand}, L.~A. 1997, \aj, 113, 1733, \dodoi{10.1086/118389}

\bibitem[{{Hillenbrand}(2021)}]{hillenbrand2021csss}
{Hillenbrand}, L.~A. 2021, in Cambridge Workshop on Cool Stars, Stellar
  Systems, and the Sun, Cambridge Workshop on Cool Stars, Stellar Systems, and
  the Sun, 280, \dodoi{10.5281/zenodo.4567863}

\bibitem[{{Jackson} {et~al.}(2020){Jackson}, {Jeffries}, {Wright}, {Rand ich},
  {Sacco}, {Pancino}, {Cantat-Gaudin}, {Gilmore}, {Vallenari}, {Bensby},
  {Bayo}, {Costado}, {Franciosini}, {Gonneau}, {Hourihane}, {Lewis}, {Monaco},
  {Morbidelli}, \& {Worley}}]{2020MNRAS.496.4701J}
{Jackson}, R.~J., {Jeffries}, R.~D., {Wright}, N.~J., {et~al.} 2020, \mnras,
  496, 4701, \dodoi{10.1093/mnras/staa1749}

\bibitem[{{Jayasinghe} {et~al.}(2019){Jayasinghe}, {Stanek}, {Kochanek},
  {Shappee}, {Holoien}, {Thompson}, {Prieto}, {Dong}, {Pawlak}, {Pejcha},
  {Shields}, {Pojmanski}, {Otero}, {Britt}, \& {Will}}]{2019MNRAS.486.1907J}
{Jayasinghe}, T., {Stanek}, K.~Z., {Kochanek}, C.~S., {et~al.} 2019, \mnras,
  486, 1907, \dodoi{10.1093/mnras/stz844}

\bibitem[{{Jeffries} {et~al.}(2011){Jeffries}, {Littlefair}, {Naylor}, \&
  {Mayne}}]{2011MNRAS.418.1948J}
{Jeffries}, R.~D., {Littlefair}, S.~P., {Naylor}, T., \& {Mayne}, N.~J. 2011,
  \mnras, 418, 1948, \dodoi{10.1111/j.1365-2966.2011.19613.x}

\bibitem[{{Jordi} {et~al.}(2005){Jordi}, {Grebel}, \&
  {Ammon}}]{2005AN....326..657J}
{Jordi}, K., {Grebel}, E.~K., \& {Ammon}, K. 2005, Astronomische Nachrichten,
  326, 657

\bibitem[{{Kenyon} \& {Hartmann}(1995)}]{1995ApJS..101..117K}
{Kenyon}, S.~J., \& {Hartmann}, L. 1995, \apjs, 101, 117,
  \dodoi{10.1086/192235}

\bibitem[{{Koenig} {et~al.}(2015){Koenig}, {Hillenbrand}, {Padgett}, \&
  {DeFelippis}}]{2015AJ....150..100K}
{Koenig}, X., {Hillenbrand}, L.~A., {Padgett}, D.~L., \& {DeFelippis}, D. 2015,
  \aj, 150, 100, \dodoi{10.1088/0004-6256/150/4/100}

\bibitem[{{Kounkel} {et~al.}(2018){Kounkel}, {Covey}, {Su{\'a}rez},
  {Rom{\'a}n-Z{\'u}{\~n}iga}, {Hernandez}, {Stassun}, {Jaehnig}, {Feigelson},
  {Pe{\~n}a Ram{\'\i}rez}, {Roman-Lopes}, {Da Rio}, {Stringfellow}, {Kim},
  {Borissova}, {Fern{\'a}ndez-Trincado}, {Burgasser},
  {Garc{\'\i}a-Hern{\'a}ndez}, {Zamora}, {Pan}, \&
  {Nitschelm}}]{2018AJ....156...84K}
{Kounkel}, M., {Covey}, K., {Su{\'a}rez}, G., {et~al.} 2018, \aj, 156, 84,
  \dodoi{10.3847/1538-3881/aad1f1}

\bibitem[{{Krumholz}(2014)}]{2014PhR...539...49K}
{Krumholz}, M.~R. 2014, \physrep, 539, 49,
  \dodoi{10.1016/j.physrep.2014.02.001}

\bibitem[{{Krumholz} \& {Tan}(2007)}]{2007ApJ...654..304K}
{Krumholz}, M.~R., \& {Tan}, J.~C. 2007, \apj, 654, 304, \dodoi{10.1086/509101}

\bibitem[{{Kuhn} {et~al.}(2019){Kuhn}, {Hillenbrand}, {Sills}, {Feigelson}, \&
  {Getman}}]{2019ApJ...870...32K}
{Kuhn}, M.~A., {Hillenbrand}, L.~A., {Sills}, A., {Feigelson}, E.~D., \&
  {Getman}, K.~V. 2019, \apj, 870, 32, \dodoi{10.3847/1538-4357/aaef8c}

\bibitem[{{Lee} {et~al.}(2015){Lee}, {Seon}, \& {Jo}}]{lee2015}
{Lee}, D., {Seon}, K.-I., \& {Jo}, Y.-S. 2015, \apj, 806, 274,
  \dodoi{10.1088/0004-637X/806/2/274}

\bibitem[{{Maddalena} {et~al.}(1986){Maddalena}, {Morris}, {Moscowitz}, \&
  {Thaddeus}}]{maddalena1986}
{Maddalena}, R.~J., {Morris}, M., {Moscowitz}, J., \& {Thaddeus}, P. 1986,
  \apj, 303, 375, \dodoi{10.1086/164083}

\bibitem[{{Mathis}(1990)}]{1990ARA&A..28...37M}
{Mathis}, J.~S. 1990, \araa, 28, 37,
  \dodoi{10.1146/annurev.aa.28.090190.000345}

\bibitem[{{Murdin} \& {Penston}(1977)}]{1977MNRAS.181..657M}
{Murdin}, P., \& {Penston}, M.~V. 1977, \mnras, 181, 657,
  \dodoi{10.1093/mnras/181.4.657}

\bibitem[{{Naylor} \& {Jeffries}(2006)}]{2006MNRAS.373.1251N}
{Naylor}, T., \& {Jeffries}, R.~D. 2006, \mnras, 373, 1251,
  \dodoi{10.1111/j.1365-2966.2006.11099.x}

\bibitem[{{Olney} {et~al.}(2020){Olney}, {Kounkel}, {Schillinger}, {Scoggins},
  {Yin}, {Howard}, {Covey}, {Hutchinson}, \& {Stassun}}]{2020AJ....159..182O}
{Olney}, R., {Kounkel}, M., {Schillinger}, C., {et~al.} 2020, \aj, 159, 182,
  \dodoi{10.3847/1538-3881/ab7a97}

\bibitem[{{Pecaut} \& {Mamajek}(2013)}]{2013ApJS..208....9P}
{Pecaut}, M.~J., \& {Mamajek}, E.~E. 2013, \apjs, 208, 9,
  \dodoi{10.1088/0067-0049/208/1/9}

\bibitem[{Pedregosa {et~al.}(2011)Pedregosa, Varoquaux, Gramfort, Michel,
  Thirion, Grisel, Blondel, Prettenhofer, Weiss, Dubourg, Vanderplas, Passos,
  Cournapeau, Brucher, Perrot, \& Duchesnay}]{scikit-learn}
Pedregosa, F., Varoquaux, G., Gramfort, A., {et~al.} 2011, Journal of Machine
  Learning Research, 12, 2825

\bibitem[{{Prisinzano} {et~al.}(2019){Prisinzano}, {Damiani}, {Kalari},
  {Jeffries}, {Bonito}, {Micela}, {Wright}, {Jackson}, {Tognelli}, {Guarcello},
  {Vink}, {Klutsch}, {Jim{\'e}nez-Esteban}, {Roccatagliata},
  {Tautvai{\v{s}}ien{\.{e}}}, {Gilmore}, {Randich}, {Alfaro}, {Flaccomio},
  {Koposov}, {Lanzafame}, {Pancino}, {Bergemann}, {Carraro}, {Franciosini},
  {Frasca}, {Gonneau}, {Hourihane}, {Jofr{\'e}}, {Lewis}, {Magrini}, {Monaco},
  {Morbidelli}, {Sacco}, {Worley}, \& {Zaggia}}]{2019A&A...623A.159P}
{Prisinzano}, L., {Damiani}, F., {Kalari}, V., {et~al.} 2019, \aap, 623, A159,
  \dodoi{10.1051/0004-6361/201834870}

\bibitem[{{Rebull} {et~al.}(2018){Rebull}, {Stauffer}, {Cody}, {Hillenbrand},
  {David}, \& {Pinsonneault}}]{2018AJ....155..196R}
{Rebull}, L.~M., {Stauffer}, J.~R., {Cody}, A.~M., {et~al.} 2018, \aj, 155,
  196, \dodoi{10.3847/1538-3881/aab605}

\bibitem[{{Rebull} {et~al.}(2016){Rebull}, {Stauffer}, {Bouvier}, {Cody},
  {Hillenbrand}, {Soderblom}, {Valenti}, {Barrado}, {Bouy}, {Ciardi},
  {Pinsonneault}, {Stassun}, {Micela}, {Aigrain}, {Vrba}, {Somers},
  {Christiansen}, {Gillen}, \& {Collier Cameron}}]{2016AJ....152..113R}
{Rebull}, L.~M., {Stauffer}, J.~R., {Bouvier}, J., {et~al.} 2016, \aj, 152,
  113, \dodoi{10.3847/0004-6256/152/5/113}

\bibitem[{{Sacco} {et~al.}(2008){Sacco}, {Franciosini}, {Randich}, \&
  {Pallavicini}}]{2008A&A...488..167S}
{Sacco}, G.~G., {Franciosini}, E., {Randich}, S., \& {Pallavicini}, R. 2008,
  \aap, 488, 167, \dodoi{10.1051/0004-6361:20079049}

\bibitem[{{Simon} {et~al.}(2019){Simon}, {Guilloteau}, {Beck}, {Chapillon}, {Di
  Folco}, {Dutrey}, {Feiden}, {Grosso}, {Pi{\'e}tu}, {Prato}, \&
  {Schaefer}}]{2019ApJ...884...42S}
{Simon}, M., {Guilloteau}, S., {Beck}, T.~L., {et~al.} 2019, \apj, 884, 42,
  \dodoi{10.3847/1538-4357/ab3e3b}

\bibitem[{{Skiff}(2014)}]{2014yCat....102023S}
{Skiff}, B.~A. 2014, VizieR Online Data Catalog, B/mk

\bibitem[{{Soderblom}(2010)}]{2010ARA&A..48..581S}
{Soderblom}, D.~R. 2010, \araa, 48, 581,
  \dodoi{10.1146/annurev-astro-081309-130806}

\bibitem[{{Somers} {et~al.}(2020){Somers}, {Cao}, \&
  {Pinsonneault}}]{2020ApJ...891...29S}
{Somers}, G., {Cao}, L., \& {Pinsonneault}, M.~H. 2020, \apj, 891, 29,
  \dodoi{10.3847/1538-4357/ab722e}

\bibitem[{{Somers} \& {Pinsonneault}(2015)}]{2015ApJ...807..174S}
{Somers}, G., \& {Pinsonneault}, M.~H. 2015, \apj, 807, 174,
  \dodoi{10.1088/0004-637X/807/2/174}

\bibitem[{{Somers} {et~al.}(2017){Somers}, {Stauffer}, {Rebull}, {Cody}, \&
  {Pinsonneault}}]{2017ApJ...850..134S}
{Somers}, G., {Stauffer}, J., {Rebull}, L., {Cody}, A.~M., \& {Pinsonneault},
  M. 2017, \apj, 850, 134, \dodoi{10.3847/1538-4357/aa93ed}

\bibitem[{{Sullivan} \& {Kraus}(2021)}]{2021ApJ...912..137S}
{Sullivan}, K., \& {Kraus}, A.~L. 2021, \apj, 912, 137,
  \dodoi{10.3847/1538-4357/abf044}

\bibitem[{{Tan} {et~al.}(2006){Tan}, {Krumholz}, \&
  {McKee}}]{2006ApJ...641L.121T}
{Tan}, J.~C., {Krumholz}, M.~R., \& {McKee}, C.~F. 2006, \apjl, 641, L121,
  \dodoi{10.1086/504150}

\bibitem[{{Zari} {et~al.}(2019){Zari}, {Brown}, \& {de
  Zeeuw}}]{2019A&A...628A.123Z}
{Zari}, E., {Brown}, A.~G.~A., \& {de Zeeuw}, P.~T. 2019, \aap, 628, A123,
  \dodoi{10.1051/0004-6361/201935781}

\end{thebibliography}
\bibliographystyle{aasjournal}

\appendix
\restartappendixnumbering

\section{Error Propagation with Colors} \label{sec:errorprop}
We define our column vectors as the residuals:
\begin{equation}
r(m_1 - m_2) = (m_1 - m_2) - (m_1 - m_2)_0 \left( \mathrm{SpT} \right).
\end{equation}
As from \citet{2011MNRAS.411..435B}, we define the Var() variance and Cov() covariance operators, and note that:
\begin{equation}
\mathrm{Cov}(aX+bY,cW+dZ) = ac \, \mathrm{Cov}(X,W) + ad \, \mathrm{Cov}(X,Z) + bc \, \mathrm{Cov}(Y,W) + bd \, \mathrm{Cov}(Y,Z).
\end{equation}

Then assuming each band is measured independently, we have the following relations:
\begin{equation}
\begin{cases} \mathrm{Cov}(r(m_1 - m_2), r(m_3 - m_4)) & = 0, \\
\mathrm{Cov}(r(m_1 - m_2), r(m_2 - m_3)) & = -\mathrm{Var}(m_2), \\
\mathrm{Cov}(r(m_1 - m_2), r(m_1 - m_2)) & = \mathrm{Var}(m_1) + \mathrm{Var}(m_2) + \left(\frac{\partial \,  (m_1 - m_2)_{0}}{\partial \, \textnormal{SpT}} \right)^2 \mathrm{Var}\left( \textnormal{SpT} \right).
\end{cases}
\end{equation}

We define the covariance matrix $\matr{C}$ with the {\it Gaia} colors (BP-G, BP-RP, and G-RP), Pan-STARRS colors (r-J, r-H, r-Ks, r-W1, r-W2, i-J, i-H, i-Ks, i-W1, i-W2), and Johnson-Cousins colors (V-J, V-H, V-Ks, V-W1, V-W2).

As a restricted example, I show a toy covariance matrix corresponding to the color set (BP-RP, V-J, r-J, r-Ks):
\begin{equation}
\matr{C} = \begin{bmatrix}
\sigma_{G_{BP}}^2 + \sigma_{G_{RP}}^2 + \sigma_{\left(G_{BP}-G_{RP}\right)_{0,\mathrm{SpT}}}^2 & 0 & 0 & 0\\
0 & \sigma_{V}^2+\sigma_{J}^2 + \sigma_{\left(V-J\right)_{0,\mathrm{SpT}}}^2 & -\sigma_{J}^2 & 0\\
0 & -\sigma_{J}^2 & \sigma_{r}^2+\sigma_{J}^2 + \sigma_{\left(r-J\right)_{0,\mathrm{SpT}}}^2 & - \sigma_{r}^2\\
0 & 0 & - \sigma_{r}^2 & \sigma_{r}^2+\sigma_{Ks}^2 + \sigma_{\left(r-Ks\right)_{0,\mathrm{SpT}}}^2\end{bmatrix}.
\end{equation}
In which $\sigma_{i}$ corresponds to the error in the photometric band $i$, and $\sigma_{\left(m_1-m_2\right)_{0,\mathrm{SpT}}}^2 = \left(\frac{\partial \,  (m_1-m_2)_{0}}{\partial \, \textnormal{SpT}} \right)^2 \sigma _{\textnormal{SpT}}^2 $.

\section{Calibration on the Pleiades \& Praesepe} \label{sec:plepra}
The Pleiades and Praesepe clusters are ideal calibration targets for our analysis---they demonstrate the accuracy and precision of our extinction estimation technique, and they calibrate {\it Gaia} colors which are homogeneous and widely available in {\lamori}. Both are old enough that disks or circumstellar material are not a concern, providing statistics on the intrinsic error of our method compared to other color-based dereddening methods. The Pleiades cluster has an age of $\sim$125 Myr with predominantly uniform extinction \citep{2007ApJ...655..233A}. Praesepe has an age of $\sim$700 Myr with indicators of low activity and negligible foreground extinction \citep{2018ApJ...863...67G,2007ApJ...655..233A}.
\subsection{Calibration of the {\it Gaia} colors to the Pleiades} \label{sec:gaiacolorsplepra}
\begin{figure}[!thb]
\centering
\includegraphics[trim={0cm 0cm 1.5cm 1cm},clip,width=\columnwidth]{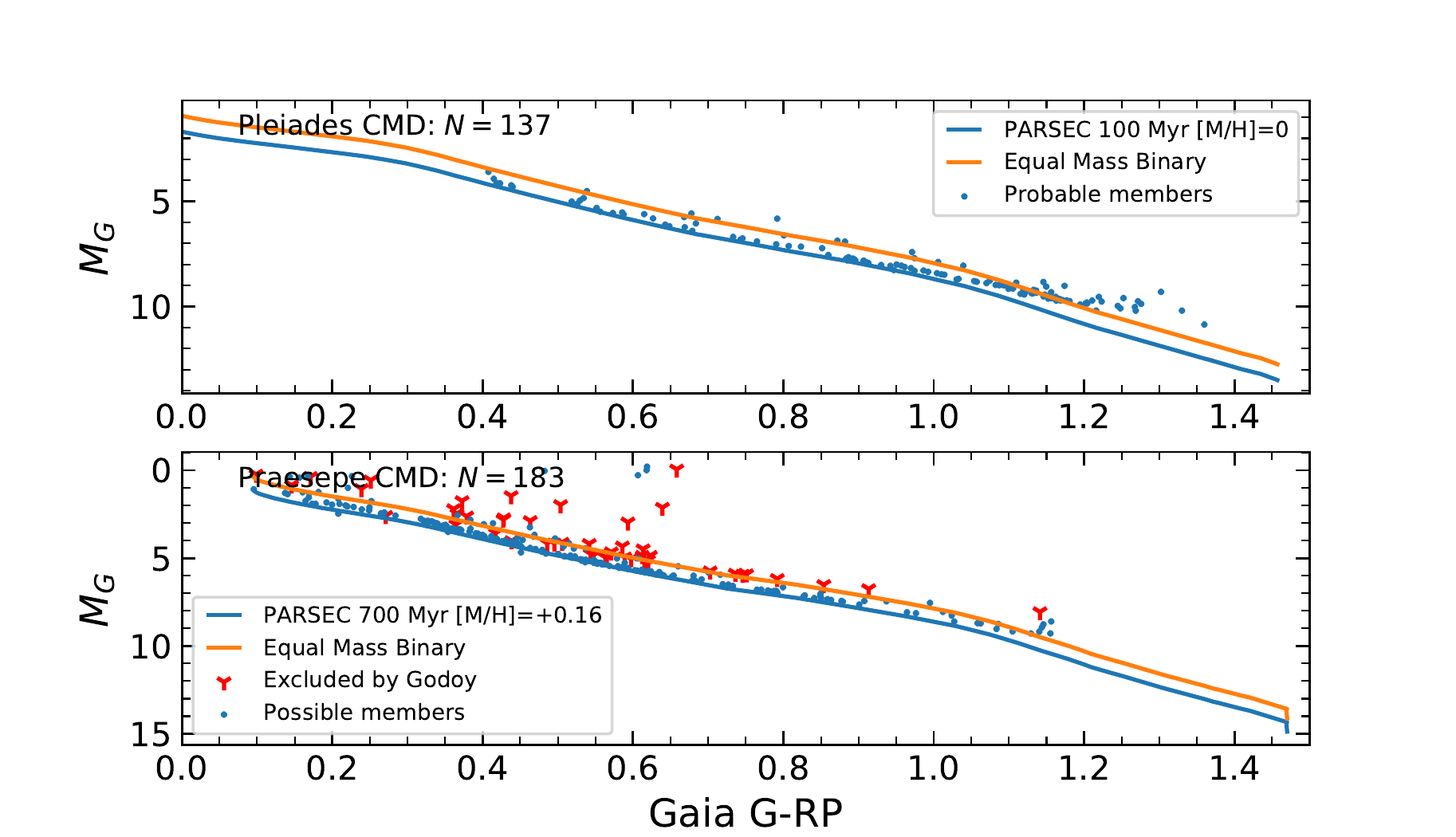}
\caption{CMD of the membership selection process for the Pleiades and Praesepe calibration sets. The noticeable drift in the PARSEC---{\it Gaia} colors in the reddest colors is part of the motivation of the calibration.} \label{fig:Figure_B1}
\end{figure}

As a calibration set, we obtain 2MASS JHK and $V-K_S$ measurements from \citet{2016AJ....152..113R}. These values of $V-K_S$ are an assortment of real V-band measurements and transformations from other bands, including SDSS $g-K_S$ and $r-K_S$ \citep{2017ApJ...850..134S}. We then perform a crossmatch to the PanSTARRS DR1 dataset \citep{2016arXiv161205560C,flewelling2016} to obtain griz data, and a crossmatch to the {\it Gaia} DR2 dataset for the {\it Gaia} photometric bands and parallax information. For Praesepe, we use the sample from \citet{2007ApJ...655..233A} crossmatched to the {\it Gaia} DR2 dataset. We obtain membership data for this cluster from \citet{2021arXiv210101183G}.
In Fig. \ref{fig:Figure_B1}, we show our selected population on a CMD alongside PARSEC---{\it Gaia} colors, showing the need to correct the theoretical colors to avoid trends in extinction.

\begin{figure}
\centering
\includegraphics[trim={0cm 0cm 1.5cm 1cm},clip,width=\columnwidth]{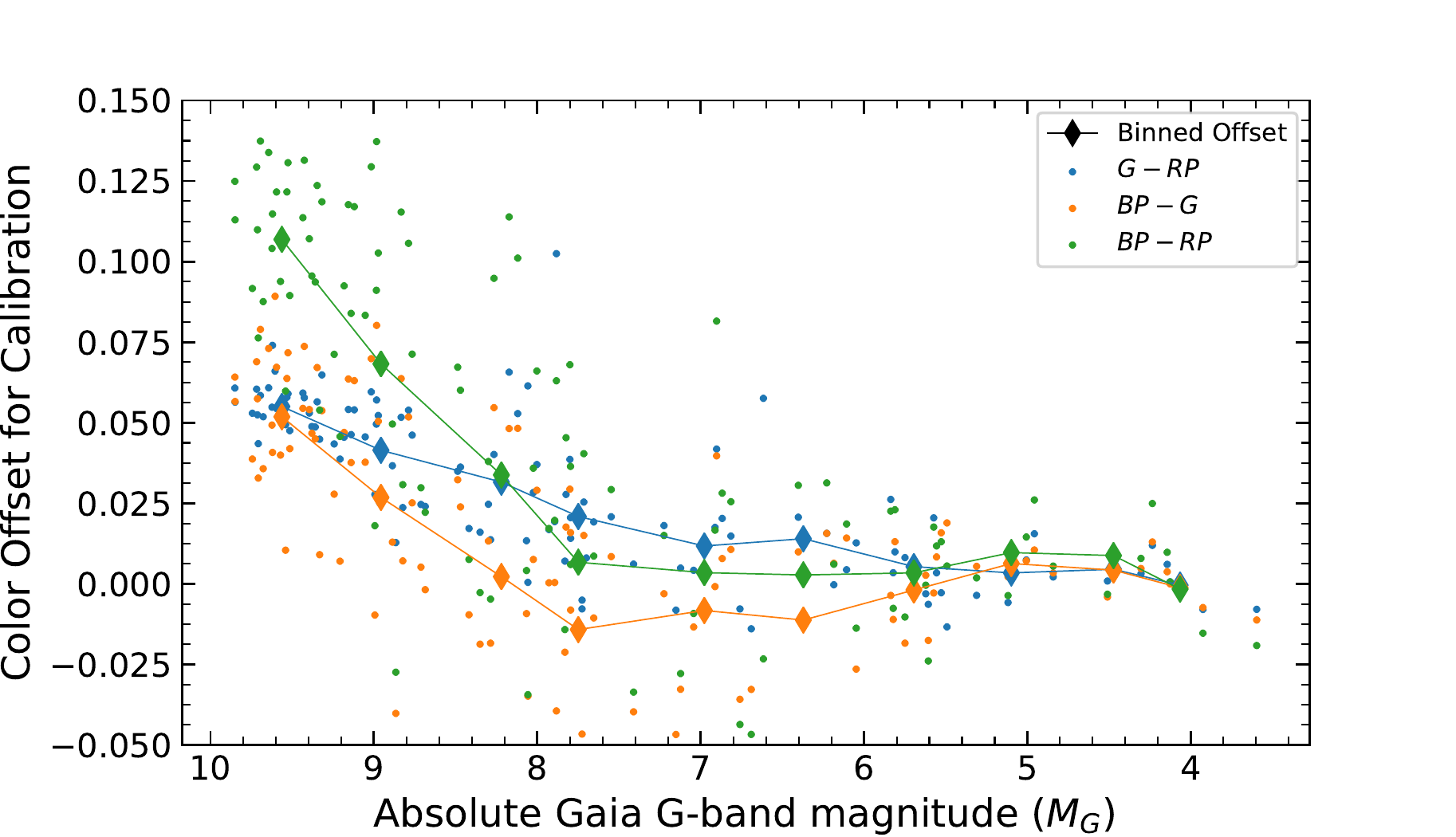}
\caption{Color calibration between dereddened ($A_V=0.12$) Pleiades {\it Gaia} colors and a PARSEC v.3.3 solar metallicity 100 Myr isochrone. We offset the PARSEC model colors with this calibration to generate intrinsic colors for {\lamori}.} \label{fig:Figure_B2}
\end{figure}

We calibrate a set of theoretical {\it Gaia} colors in PARSEC to the dereddened colors of the Pleiades with a cluster mean extinction set to $0.12$ mag \citep{2007ApJ...655..233A}. In Fig. \ref{fig:Figure_B2}, we show the difference in color between our observed data and the theoretical PARSEC {\it Gaia} colors G-RP, BP-G, and BP-RP. We plot this against absolute {\it Gaia} G magnitude for the Pleiades. This color offset is maximized for the faintest stars and ranges from 0.1 mag at the faint end of our calibration to a more typical value of 0.025 mag and below. We apply these offsets to the 5 Myr PARSEC v.3.3 {\it Gaia} colors in our analysis for the SED method in our Class III sources.

\subsection{$A_V$ SED Testing with Pleiades and Praesepe}
In these open clusters, it is possible to set the temperatures of these stars by fixing their $A_V$, using reliable colors in the optical and near-infrared as in \citet{2003A&A...401..781F}, defining as the color excess the difference between the observed color and the intrinsic color from PM13 young (Pleiades) and dwarf (Praesepe) tables:

\begin{equation}
A_V = \frac{E \left( V - K_S \right)}{0.89} = \frac{E \left( V - H \right)}{0.83} = \frac{E \left( V - J \right)}{0.73}
\end{equation}

Because of the strong temperature-extinction degeneracy, we choose a temperature for each star that provides a mean stellar extinction as close as possible to the mean value from \citet{2007ApJ...655..233A}. We also characterize the temperature uncertainty by inferring a temperature for each color assuming the same extinction, and measuring the dispersion between the temperature estimates for each individual star. This leads to an internal error of roughly $10$---$20$ K for the Pleiades and Praesepe temperatures.

\begin{figure}
\centering
\includegraphics[trim={0cm 0cm 0cm 0cm},clip,width=\columnwidth]{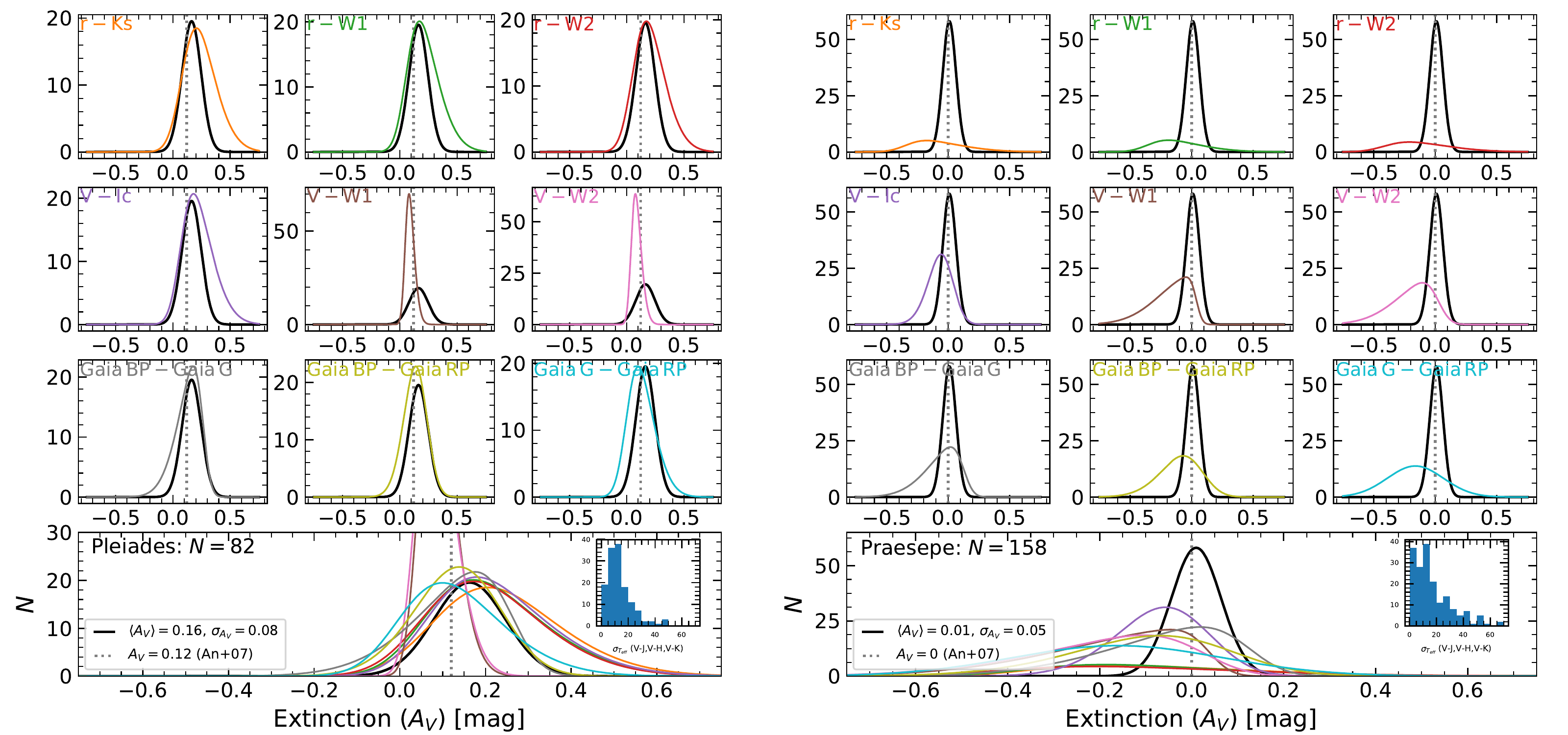}
\caption{Comparison of individual stellar extinction values for the Pleiades and Praesepe between our SED-fitting method (black) and single-color extinctions (assorted colors). Top nine figures: comparisons of individual colors against the SED-fitting solution. Dashed lines: $A_V$ values from \citet{2007ApJ...655..233A}. Inset plots: measured temperature dispersion as a function of the colors used to infer temperatures (V-J, V-H, V-K).} \label{fig:Figure_B3}
\end{figure}

We then test our aggregate $A_V$ SED method on the Pleiades and Praesepe, using the same color vector as used in {\lamOri} but ignoring the colors (V-J, V-H, V-Ks) that we used to set the temperature. 
The intrinsic colors for the Pleiades included the base PM13 young color table, the \citet{2017AJ....153..188F} theoretical colors (ugriz), and the calibrated PARSEC $100$ Myr v.3.3 solar metallicity isochrone ({\it Gaia}). For the considerably older Praesepe cluster, we use the \citet{2013ApJS..208....9P} dwarf color table augmented with the \citet{2005AN....326..657J} SDSS empirical color transformation (ugriz), and an unmodified PARSEC $700$ Myr v.3.3 [Fe/H] = 0.16 isochrone ({\it Gaia}). In Fig. \ref{fig:Figure_B3}, we show that the derived $A_V$ with our SED-based approach obtains a consistent result with \citet{2007ApJ...655..233A}, featuring a low star-by-star variation in $A_V$.

\end{document}